\UseRawInputEncoding

\documentclass[sigconf]{acmart}

\AtBeginDocument{%
  }

\copyrightyear{2025}
\acmYear{2025}
\setcopyright{acmlicensed}
\acmConference[MICRO '25]{58th IEEE/ACM International Symposium on Microarchitecture}{October 18--22, 2025}{Seoul, Republic of Korea}
\acmBooktitle{58th IEEE/ACM International Symposium on Microarchitecture (MICRO '25), October 18--22, 2025, Seoul, Republic of Korea}
\acmDOI{10.1145/3725843.3756099}
\acmISBN{979-8-4007-1573-0/2025/10}

\newcommand\mynuma[1]{\ifcase#1 \or \ding{172}\or \ding{173}\or
  \ding{174}\or \ding{175}\or \ding{176}\or \ding{177}%
  \or \ding{178}\or \ding{179}\or \ding{180}\or \ding{181}\else *\fi\relax}
\newcommand\mynumb [1]{\ifcase#1 \or \ding{182}\or \ding{183}\or
  \ding{184}\or \ding{185}\or \ding{186}\or \ding{187}%
  \or \ding{188}\or \ding{189}\or \ding{190}\or \ding{191}\else *\fi\relax}

\definecolor{assign_color}{rgb}{0.0, 0.0, 1.0}

\definecolor{note_color}{rgb}{1.0, 0.0, 0.0}
\newcommand{\TODO}[1]{\textcolor{note_color}{#1}}

\definecolor{ra}{HTML}{6D8764}
\definecolor{rb}{HTML}{C3ABD0}
\definecolor{rc}{HTML}{F0A30A}
\definecolor{rd}{HTML}{1BA1E2}
\definecolor{re}{HTML}{A0522D}
\definecolor{rf}{HTML}{0000FF}
\definecolor{rose}{HTML}{D36C8C}

\newcommand{\name}{RTGS}
\usepackage{tikz}
\newcommand{\lefthalfblack}{
  \tikz[baseline=-0.6ex]{
    \draw[thick] (0,0) circle (0.9ex);
    \fill[black] (0,0) -- (90:0.9ex) arc (90:270:0.9ex) -- cycle;
  }
}

\usepackage{hyperxmp} 
\usepackage{enumitem}
\usepackage[textsize=footnotesize,textwidth=1.3\linewidth]{todonotes}
\usepackage{mathtools}
\usepackage{amsmath}
\usepackage{multirow} 
\usepackage{makecell}
\usepackage{subcaption}
\usepackage{soul}
\usepackage{enumitem}
\usepackage{algorithm}
\usepackage{colortbl}
\usepackage{listings}
\usepackage{pifont}
\usepackage{float}

\definecolor{mygreen}{HTML}{006400} 
\definecolor{myred}{HTML}{EA4436} 

\newcommand{\cmark}{\textcolor{green!60!black}{\textbf{\ding{51}}}} 
\newcommand{\xmark}{\textcolor{red}{\textbf{\ding{55}}}}     

\begin{document}

\newpage
\setcounter{page}{1}

\title{\name: Real-Time 3D Gaussian Splatting SLAM via Multi-Level Redundancy Reduction}

\newcommand{\affnum}[1]{\textsuperscript{#1}}
\newcommand{\cofirst}{\textsuperscript{*}}

\author{%
{
Leshu Li\cofirst\textsuperscript{1},
Jiayin Qin\cofirst\textsuperscript{1},
Jie Peng\textsuperscript{2},
Zishen Wan\textsuperscript{3},
Huaizhi Qu\textsuperscript{2},
Ye Han\textsuperscript{1},
Pingqing Zheng\textsuperscript{1},
Hongsen Zhang\textsuperscript{1},
Yu (Kevin) Cao\textsuperscript{1},
Tianlong Chen\textsuperscript{2},
Yang (Katie) Zhao\textsuperscript{1}
}
}

\affiliation{%
  \textsuperscript{1}Department of Electrical and Computer Engineering, University of Minnesota, Twin Cities 
  \country{USA}\\
  \textsuperscript{2}Department of Computer Science, University of North Carolina at Chapel Hill \country{USA}\\
  \textsuperscript{3}Department of Electrical and Computer Engineering, Georgia Institute of Technology \country{USA}
}
\email{{li003385, qin00162, yangzhao}@umn.edu}

\begin{abstract}

3D Gaussian Splatting (3DGS) based Simultaneous Localization and Mapping (SLAM) systems can largely benefit from 3DGS's state-of-the-art rendering efficiency and accuracy, but have not yet been adopted in resource-constrained edge devices due to insufficient speed.
Addressing this, we identify notable redundancies across the SLAM pipeline for acceleration. While conceptually straightforward, practical approaches are required to minimize the overhead associated with identifying and eliminating these redundancies.

In response, we propose {\name}, an algorithm-hardware co-design framework that comprehensively reduces the redundancies for real-time 3DGS-SLAM on edge. 
To minimize the overhead, {\name} fully leverages the characteristics of the 3DGS-SLAM pipeline. 

On the algorithm side, we introduce (1) an adaptive Gaussian pruning step to remove the redundant Gaussians by reusing gradients computed during backpropagation; and (2) a dynamic downsampling technique that directly reuses the keyframe identification and alpha computing steps to eliminate redundant pixels. On the hardware side, we propose (1) a subtile-level streaming strategy and a pixel-level pairwise scheduling strategy that mitigates workload imbalance via a Workload Scheduling Unit (WSU) guided by previous iteration information; (2) a Rendering and Backpropagation (R\&B) Buffer that accelerates the rendering backpropagation by reusing intermediate data computed during rendering; and (3) a Gradient Merging Unit (GMU) to reduce intensive memory accesses caused by atomic operations while enabling pipelined aggregation. 

Integrated into an edge GPU, {\name} achieves real-time performance ($\geq$30 FPS) on four datasets and three algorithms, with up to 82.5$\times$ energy efficiency over the baseline and negligible quality loss. 
Code is available at \underline{\href{https://github.com/UMN-ZhaoLab/RTGS}{https://github.com/UMN-ZhaoLab/RTGS}}.

\vspace{-0.8 em}
\end{abstract}


\maketitle
\pagestyle{plain}

\begingroup
\renewcommand{\thefootnote}{\fnsymbol{footnote}} 
\setcounter{footnote}{1}                          
\footnotetext{Both authors contributed equally to this research.}
\endgroup


\section{Introduction}
\label{sec:intro}

The growing demand for efficient and high-fidelity Simultaneous Localization and Mapping (SLAM) systems in augmented/virtual reality (AR/VR), autonomous driving, and robotic navigation has driven the adoption of various 3D representations over the years. Classical SLAM algorithms with compute-intensive scene representations, like polygonal meshes~\cite{mesh,mesh1,mesh2,mesh3}, point clouds~\cite{pointcloud,pointcloud1,pointcloud2,pointcloud3}, or voxels~\cite{voxel,voxel1,voxel2,voxel3}, are often challenged by inadequate quality and lack of real-time performance. NeRF (Neural Radiance Fields)~\cite{nerf}, on the other hand, is known for its memory efficiency and photorealistic quality but it suffers from slow rendering speed~\cite{niceslam,nerf_survey,nerf_survey2}. 
Recent advances have focused on more efficient representations of 3D scenes. One promising approach is 3D Gaussian Splatting (3DGS), which explicitly represents 3D scenes with ellipsoidal 3D Gaussians~\cite{3DGS}. Applying 3DGS to SLAM, termed 3DGS-SLAM, offers several advantages, including faster and more photorealistic rendering, the flexibility to increase map capacity, full utilization of dense photometric losses, and direct gradient backpropagation to parameters to facilitate fast training~\cite{photoslam,GSSLAM,gaussianslam}.

Despite these advantages, 3DGS-SLAM still struggles to achieve real-time throughput, i.e., 30 FPS, especially on edge devices~\cite{ARVR, ARVR2}. For example, on a state-of-the-art (SOTA) edge GPU~\cite{ONX}, even the most efficient 3DGS-SLAM algorithms run at or below 15 FPS for tracking, let alone for the overall tracking and mapping SLAM pipeline. This gap hinders the wide adoption of edge devices that benefit from 3DGS-SLAM solutions. 
To bridge this gap, we first conduct extensive profiling measurements of the SOTA 3DGS-SLAM solutions and identify significant redundancies across the SLAM pipeline that can be leveraged for acceleration. While existing methods have explored some of these redundancies (see Tab.~\ref{tab:3dgs_methods_comparison}), a comprehensive exploration across all redundancy sources is essential to enable real-time 3DGS-SLAM on edge devices.

\begin{table}[t]
    \centering
    \renewcommand{\arraystretch}{1.2} 
    \setlength{\tabcolsep}{5pt} 
    \caption{Comparison of {\name} with prior 3DGS solutions.}
    \vspace{-10pt}
    \label{tab:3dgs_methods_comparison}
    \resizebox{\columnwidth}{!}{%
    \begin{tabular}{c|c||c|c|c|c|c|c}
        \toprule
        \multicolumn{2}{c||}{\textbf{Method}} & \textbf{RTGS} & \makecell{\textbf{GauSPU}\\~\cite{GauSPU} }& \makecell{\textbf{GSArch}\\~\cite{gsarch}} & \makecell{\textbf{MetaSapiens}\\~\cite{lin2025metasapiens}} & \makecell{\textbf{Taming 3DGS}\\~\cite{Taming}} & \makecell{\textbf{DISTWAR}\\~\cite{distwar}} \\
        \midrule
        \multirow{5}{*}{\makecell{Supported \\ Functions}} 
        & Inference & \cmark & \cmark & \cmark & \cmark & \xmark & \xmark \\
        \cline{2-8}
        & Training & \cmark & \cmark & \cmark & \xmark$^4$ & \cmark & \cmark \\
        \cline{2-8} 
        & \makecell{SLAM \\Tracking}
            & \cmark  
            & \cmark  
            & \lefthalfblack$3$ 
            & \xmark$^4$  
            & \xmark\hspace{0.3em}$^5$ 
            & \lefthalfblack$6$ \\
         \cline{2-8}
        & \makecell{SLAM\\ Mapping} & \cmark & \lefthalfblack$^1$ & \lefthalfblack$^3$ & \xmark$^4$ & \xmark\hspace{0.3em}$^5$ & \lefthalfblack$^6$\\
         \midrule
        \multirow{7}{*}{\makecell{Utilized \\ Redundancy}} 
        & Gaussians & \cmark & \xmark & \xmark$^3$ & \xmark\hspace{0.3em}$^4$ & \cmark$^5$ & \xmark \\
        \cline{2-8}
        & Pixels & \cmark & \cmark$^1$ & \xmark & \xmark & \xmark & \xmark \\
         \cline{2-8}
        & \makecell{Comp. in \\Blending BP.}  & \cmark & \xmark & \xmark & \xmark & \xmark & \xmark \\
        \cline{2-8}
        & \makecell{Grad. Agg. \\Mem. Accesses} & \cmark & \cmark & \cmark & \xmark & \xmark & \cmark$^6$ \\
        \cline{2-8}
        & \makecell{Imbalanced \\Workload} & \cmark & \cmark\hspace{0.3em}$^2$ & \xmark & \cmark\hspace{0.3em}$^2$ & \xmark & \xmark \\
        \bottomrule
    \end{tabular}%
    }
        \raggedright
    \footnotesize 
    $^1$ 
    GauSPU~\cite{GauSPU} identifies redundant pixels by counting the number of Gaussians involved during SLAM tracking stages, where the Gaussians remain fixed. However, SLAM mapping stages continuously introduce new Gaussians.\\
    $^2$
    GauSPU~\cite{GauSPU} and MetaSapien~\cite{lin2025metasapiens} address tile-level workload imbalance through streaming and tile merging, but they ignore pixel-level workload imbalance.\\
    $^3$
    One of the key ideas of GSArch~\cite{gsarch} is to determine Gaussian importance by the total number of pixels a Gaussian covers across all rendering frames, but this can harm SLAM performance by ignoring Gaussians critical for tracking in the current frame.\\
    $^4$
    MetaSapiens~\cite{lin2025metasapiens} prunes Gaussians for inference with prohibitive training overhead. \\
    
    $^5$
    Taming 3DGS~\cite{Taming} relies on gradient changes in the first 500 iterations 
    to predict important Gaussians and requires thousands of iterations to gradually converge, which is inefficient for 3DGS-SLAMs with only 15-100 iterations per frame.\\
    $^6$ 
    DISTWAR~\cite{distwar} uses warp-level gradient merging to reduce redundant atomic memory accesses in Gaussian gradient accumulation, with a warp as the smallest compute unit. The sparsity of Gaussians in SLAMs limits its effectiveness. \\ 
\vspace{-14pt}
\end{table}


Although redundancy reduction is conceptually straightforward, the overhead of identifying and eliminating redundancies remains a critical challenge, potentially negating the achieved improvements. 
To this end, we propose {\name}, an algorithm-hardware co-designed framework, to reduce redundancies for real-time 3DGS-SLAM comprehensively. 
The key novelty of {\name} lies in its ability to leverage the existing 3DGS-SLAM pipeline to manage the overhead of identifying and reducing redundancies. Our contributions are as follows:


\begin{itemize}[leftmargin=*]
\item We conduct a comprehensive analysis of various 3DGS-SLAM methods and identify significant multi-level redundancies across the 3DGS-SLAM pipeline for acceleration.

\item On the \textbf{algorithm} side, we introduce (1) an adaptive Gaussian pruning step to remove redundant Gaussians by reusing gradients computed during backpropagation. In addition, leveraging our discovery that non-keyframes contain a large amount of redundant pixels, we propose (2) a dynamic downsampling technique that directly reuses the keyframe identification and alpha computing steps to eliminate redundant pixels.  

 \item On the \textbf{hardware} side, we propose an edge GPU integrated plug-in to support 3DGS-SLAM applications in real time. Our {\name} features: (1) two complementary techniques to reduce workload imbalance, including a subtile-level streaming technique and a pixel-level scheduling technique. These are efficiently realized in the Workload Scheduling Unit (WSU), which exploits inter-iteration similarity to reuse scheduling patterns with minimal overhead, (2) a Rendering and Backpropagation (R\&B) Buffer to store reusable parameters between rendering and backpropagation with negligible memory overhead, and
 (3) a pipelined Gradient Merging Unit (GMU) that aggregates gradients of the same address to reduce memory collisions of atomic operations.

 \item We comprehensively evaluate our design across three 3DGS-SLAMs featuring distinct pipelines on four datasets. The experiment results show that~{\name} provides up to 48.8$\times$ speedup and an 82.5$\times$ improvement in energy efficiency over the edge GPU.
\end{itemize}

\section{Background}
\label{sec:bg}

\begin{figure*}[t]
    \centering
    \begin{minipage}[b]{0.68\textwidth} 
        \centering
        \includegraphics[width=1\textwidth]{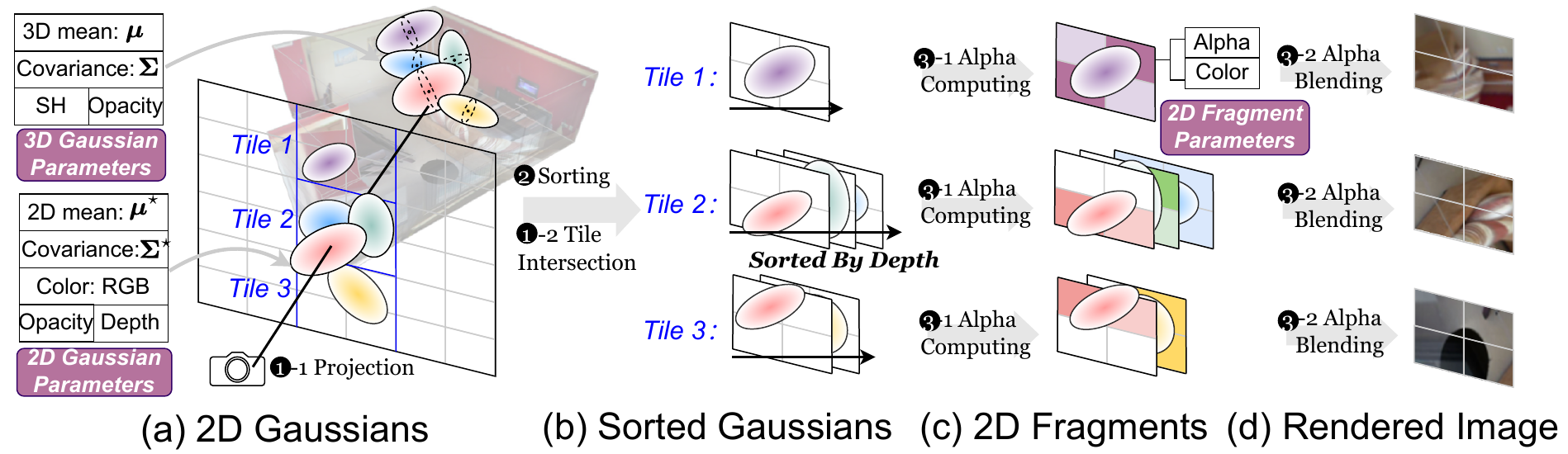} 
        \vspace{-20pt}
        \caption{Rendering pipeline: (a) Projecting 3D Gaussians into 2D Gaussians. (b) Sorting 2D Gaussians by depth. (c) Calculating the influence of each Gaussian on the pixels.} 
        \label{fig:rendering_pipline} 
    \end{minipage}
    \hfill 
    \begin{minipage}[b]{0.3\textwidth} 
        \centering
        \includegraphics[width=1\textwidth]{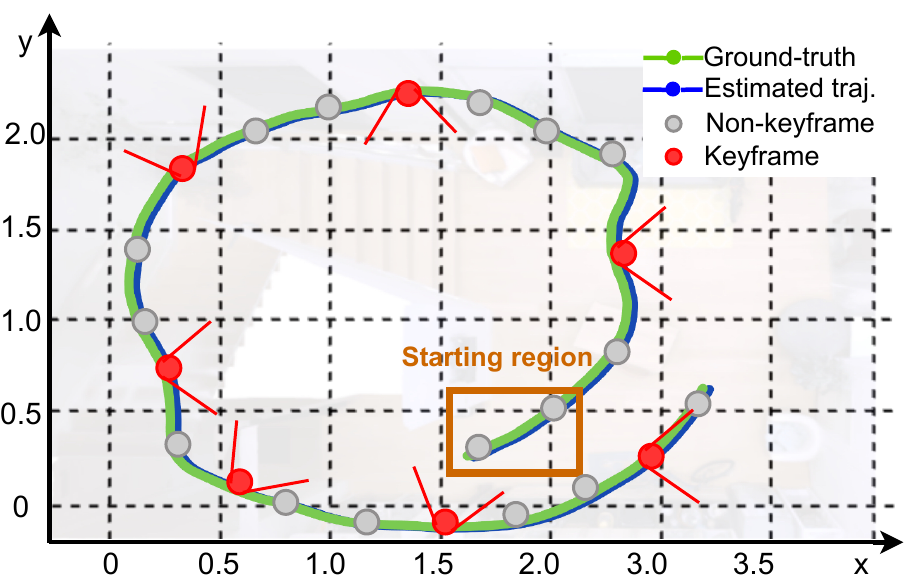} 
        \vspace{-20pt}
        \caption{Illustration of keyframes and non-keyframes in SLAM.} %
        \label{fig:SLAM_overv} 
    \end{minipage}
\vspace{-3pt}
\end{figure*}

\subsection{Preliminaries of 3D Gaussian Splatting}
\label{sec:3DGS_bg}


\textbf{Scene Representation using 3D Gaussians.} 
3DGS represents a scene using a set of ellipsoidal 3D Gaussians, denoted as ${G}$. Each 3D Gaussian is associated with trainable parameters to describe its attributes, including the 3D position mean ${\mu}$, the 3D covariance matrix ${\Sigma}$, opacity ${o}$, and color distribution ${sh}$, where $k$ is its ID .
\begin{align} 
\begin{split}
{G} &= \{G_k^{3D}:(\mu_k,\Sigma_k,o_k,sh_k)\}  \\
where\; G_k^{3D}({x}) &= \exp ( -\frac{1}{2} ({x} - {\mu}_k)^\top {\Sigma}^{-1}_k ({x} - {\mu}_k))
\label{eq:3DGS}
\end{split}
\end{align}


\noindent
\textbf{The Rendering Pipeline of 3DGS.} 
Given the 3D Gaussians and the camera pose, Fig.~\ref{fig:rendering_pipline} illustrates how the 3D Gaussians are rendered into a 2D RGB image through the following three steps: 

\textbf{Step}~\ding{182}~\textbf{Preprocessing}: This step contains two sub-steps. \textbf{Step}~\ding{182}-1~\textit{Projection} projects each ellipsoidal 3D Gaussian into an elliptical 2D Gaussian on the image plane using camera pose, resulting in 2D Gaussian attributes, e.g., 2D position ${\mu_k}^\star$, 2D covariance ${\Sigma_k}^\star$, color ${C_k}$, opacity ${o_k}$ and depth ${d_k}$. \textbf{Step}~\ding{182}-2~\textit{Tile intersection} assigns the projected 2D Gaussians to different tiles\footnote{A tile refers to a grid of pixels (e.g., 16$\times$16) to partition the image for parallel computation, following the conventional tile-based rendering implementation on GPUs.} based on their positions. 


\textbf{Step}~\ding{183}~\textbf{Sorting}:  For each pixel $P$, all covering 2D Gaussians are projected to generate \textit{fragments}, where each fragment $f_{P,k}$ denotes the contribution of the Gaussian $k$ to pixel $P$. These fragments are then sorted by depth (in forward) to ensure correct occlusion.

\textbf{Step}~\ding{184}~\textbf{Rendering}: Different from the previous two steps conducted per Gaussian, this step is performed per pixel with the basic compute unit of a 2D fragment. Specifically, a 2D fragment is a pair of one pixel and one 2D Gaussian covering it; note that one pixel may have multiple fragments since multiple 2D Gaussians may cover it. First, we compute the alpha value $\alpha_{P,k}$ for each 2D fragment in \textbf{Step}~\ding{184}-1~\textbf{Alpha Computing}. Then, in \textbf{Step}~\ding{184}-2~\textbf{Alpha Blending}, we blend the alpha values of all 2D Gaussians covering each pixel, i.e., all fragments of this pixel, to obtain the final color of each pixel.  
The alpha value $\alpha_{P,k}$ of Gaussian $G_k^{2D}$ at pixel $P$ is:
\begin{align}
\alpha_{P,k} = o_k G_k^{2D} = o_k \exp\left( -\tfrac{1}{2} (P - \boldsymbol{\mu}^\star_k)^\top \left( \boldsymbol{\Sigma}_k^\star \right)^{-1} (P - \boldsymbol{\mu}^\star_k) \right), \label{eq:alpha}
\end{align}
The per-Gaussian color contribution $\hat{\mathbf{C}}_{P,k}$ and pixel color $\mathbf{C}_P$ are:
\begin{align}
\hat{\mathbf{C}}_{P,k} = T_{P,k} \alpha_{P,k} \mathbf{C}_k, \quad \mathbf{C}_P = \sum \hat{\mathbf{C}}_{P,k}, \label{eq:color}
\end{align}
where $T_{P,k} = \prod_{n=1}^{k-1} \left( 1 - \alpha_{P,n} \right)$ represents the accumulated transparency. Note that when the transparency falls below a threshold, indicating a full occlusion for Gaussians behind, the ray rendering process can be terminated early, preserving the sequential processing order of ${C}_P$ during rendering.

\subsection{3D Gaussian Splatting-based SLAM}
\label{sec:3DGS-SLAM_bg}

\textbf{3DGS-SLAM Pipeline.}
Similar to other SLAM methods, 3DGS-SLAM is divided into two stages: tracking and mapping~\cite{SLAM1, SLAM_survey}.  
The tracking stage updates the camera pose, while the mapping stage updates the model that represents the scene, which, in the case of 3DGS-SLAM, refers to the 3D Gaussians.  
In addition, 3DGS-SLAM also distinguishes between keyframes and non-keyframes: tracking and mapping are performed on keyframes, while only tracking is conducted on non-keyframes.
The alternating process of keyframes and non-keyframes in SLAM is illustrated in Fig.\ref{fig:SLAM_overv}.

Both tracking and mapping stages involve rendering and backpropagation.
The rendering process is composed of three stages, as outlined in Sec.~\ref{sec:3DGS_bg}. However, 
during backpropagation, the sorting can be omitted by reusing the information in the forward pass.
Therefore, backpropagation only involves two steps: \textbf{Step}~\ding{185}\ \textbf{Rendering Backpropagation (BP)} and \textbf{Step}~\ding{186}\ \textbf{Preprocessing BP}.

\textbf{Step}~\ding{185}~\textbf{Rendering BP}: This step propagates the pixel color loss (i.e., $L$) to the corresponding pixel-level 2D Gaussian gradients (i.e., $dL/dG^{2D}_k[i][j]$, where $k$ is the Gaussian ID and $[i][j]$ denotes the pixel $P$ location). This includes gradients with respect to $\hat{\mathbf{C}}_{P,k}$, $a_{P,k}$, $\boldsymbol{\mu}^\star_{P,k}$, and $\boldsymbol{\Sigma}^\star_{P,k}$. Each GPU thread is responsible for computing gradients for multiple fragments (which correspond to multiple 2D Gaussians) of a single pixel.  
The pixel-level gradients $dL/dG^{2D}_k[i][j]$ are aggregated to tile-level gradients (i.e., $dL/dG_{2D}^k[m]$, where $[m]$ is the tile ID), and finally to Gaussian-level gradients $dL/dG^{2D}_k$ via atomic add operations on the GPU, which may increase redundant memory access conflicts and stalls.

It is important to highlight that rendering BP is structurally the inverse of the forward compositing process.  
The most critical step in this process is computing the gradient of the loss $\mathcal{L}$ for the opacity $\alpha_{P,k}$ of the $k$-th Gaussian along a ray, given by:
\begin{align}
\label{eq:back}
\frac{\partial \mathcal{L}}{\partial \alpha_{P,k}} = \left( \mathbf{C}_k - \sum_{n > k} \hat{\mathbf{C}}_{P,n} \right) \cdot \frac{\partial \mathcal{L}}{\partial \mathbf{C}_P}.
\end{align}


The accumulated transmittance $T_{P,i}$ is updated recursively as:
\begin{align}
\label{eq:back_2}
T_{P,k} = \frac{T}{1 - \alpha_{P,k}}.
\end{align}



\textbf{Step}~\ding{186}~\textbf{Preprocessing BP}: In this step, Gaussian-level gradients $dL/dG^{2D}_k$ are propagated back to the 3D Gaussian gradients $dL/dG^{3D}_k$.  In the mapping stage, these gradients are used to update 3D Gaussian parameters. 
The tracking process involves an additional step called Camera pose optimization, in which the gradients $dL/dP_{k}$ computed for all 3D Gaussians from their respective $dL/dG^{3D}_k$ are further aggregated to obtain the final gradient $dL/dP$, which is then used to optimize the camera pose.


\subsection{State-of-the-Art 3DGS-SLAMs}  
Thanks to the fully rasterized rendering pipeline, one of the most notable advantages of 3DGS-SLAM is its fast rendering speed and high rendering quality~\cite{rendering_Speed, rendering_Speed2}.  
However, the overall runtime performance of 3DGS-SLAM is hindered by the need for multiple training iterations per frame.  
SplaTAM~\cite{splatam} is a representation algorithm that performs both tracking and mapping for every frame, resulting in 0.78 FPS on the SOTA ONX edge GPU~\cite{ONX}. 
To improve efficiency, SOTA approaches adopt a keyframe-based mapping strategy. GS-SLAM~\cite{GSSLAM} updates the submap only at keyframes and achieves 2.34 FPS on the ONX. MonoGS~\cite{MONOGS} further extends this keyframe-based mapping design by demonstrating strong adaptability to monocular (i.e., RGB) datasets. To enhance reconstruction completeness and detail recovery in monocular scenes, MonoGS typically utilizes a larger number of Gaussians for mapping. Photo-SLAM~\cite{photoslam} adopts a hybrid design that combines traditional geometric SLAM components. Unlike the aforementioned fully end-to-end learnable approaches, Photo-SLAM relies entirely on classical geometric optimization for its tracking BP to improve tracking throughput.

Despite these advancements, the runtime performance of existing 3DGS-SLAM solutions remains significantly below the threshold required for real-time SLAM applications (\(\geq\) 30 FPS)~\cite{ARVR, ARVR2}. Tab.~\ref{tab:slam_comparison} summarizes the accuracy, speed, and storage efficiency of the four aforementioned 3DGS-SLAM algorithms.

\begin{table}[t]
    \centering
    \setlength{\tabcolsep}{2pt}
    \caption{Performance comparison of different SLAM algorithms on ONX edge GPU~\cite{ONX} using Repilica dataset ~\cite{Replica}, where Absolute Trajectory Error (ATE) measures tracking accuracy (lower is better) and Peak Signal-to-Noise Ratio (PSNR) reflects rendering fidelity (higher is better).}
    \vspace{-10pt}
    \renewcommand{\arraystretch}{1.0}
    \resizebox{\columnwidth}{!}{
    \begin{tabular}{
            >{\centering\arraybackslash}m{2.4cm}|| 
    >{\centering\arraybackslash}m{1.8cm}|
    >{\centering\arraybackslash}m{1.8cm}||
    >{\centering\arraybackslash}m{1.8cm}|
    >{\centering\arraybackslash}m{1.8cm}||
    >{\centering\arraybackslash}m{2.3cm}|
    >{\centering\arraybackslash}m{1.5cm}
    }
        \toprule
        \multirow{2}{*}{\textbf{Algorithm}} & 
        \multicolumn{2}{c||}{\makecell{\textbf{Accuracy}\\\textbf{Performance}}} & 
        \multicolumn{2}{c||}{\makecell{\textbf{Speed}\\\textbf{Performance}}} & 
        \multicolumn{1}{c|}{\makecell{\textbf{Storage}\\\textbf{Efficiency}}} & 
        \makecell{\textbf{Dataset}} \\
        \cline{2-7}
        & \makecell{ATE \\(cm) $\downarrow$} & \makecell{PSNR\\(dB) $\uparrow$} & \makecell{Tracking\\FPS $\uparrow$} & \makecell{Overall\\FPS$^1$ $\uparrow$} & \makecell{Peak Gaussian\\Mem. Capacity $\downarrow$} & \makecell{Monocular\\Support} \\
        \midrule

        \makecell{SplaTAM~\cite{splatam}} 
        & \cellcolor[HTML]{FFE599}Medium & \cellcolor[HTML]{B6D7A8}High 
        & \cellcolor[HTML]{EA9999}Slow & \cellcolor[HTML]{EA9999}Slow 
        & \cellcolor[HTML]{EA9999}Inefficient & \makecell{\Large \xmark} \\
        & (0.36-2.25) & (25.12-34.11) & (0.26-0.46) & (0.42-0.78) & (7-10 GB) & \\

        \midrule
        \makecell{GS-SLAM~\cite{GSSLAM}} 
        & \cellcolor[HTML]{EA9999}Low & \cellcolor[HTML]{B6D7A8}High
        & \cellcolor[HTML]{FFE599}Moderate & \cellcolor[HTML]{FFE599}Moderate 
        & \cellcolor[HTML]{EA9999}Inefficient & \makecell{\Large \xmark} \\
        & (0.5-3.7) & (21.6-31.56) & (1.45-2.37) & (1.45-2.34) & (8-12 GB) & \\

        \midrule
        \makecell{MonoGS~\cite{MONOGS}} 
        & \cellcolor[HTML]{B6D7A8}High & \cellcolor[HTML]{B6D7A8}High
        & \cellcolor[HTML]{FFE599}Moderate & \cellcolor[HTML]{FFE599}Moderate 
        & \cellcolor[HTML]{EA9999}Inefficient & \makecell{\Large \cmark} \\
        & (0.32-1.58) & (25.82-34.83) & (0.81-1.32) & (0.83-1.3) & (13-15 GB) & \\

        \midrule
        \makecell{Photo-SLAM~\cite{photoslam}}  
        & \cellcolor[HTML]{EA9999}Low & \cellcolor[HTML]{B6D7A8}High
        & \cellcolor[HTML]{B6D7A8}Fast$^2$ & \cellcolor[HTML]{B6D7A8}Fast$^2$ 
        & \cellcolor[HTML]{B6D7A8}Acceptable & \makecell{\Large \cmark} \\
        & (0.53-2.8) & (20.12-31.97) & (11.7-14.3) & (8.3-9.4) & (4-5 GB) & \\

        \bottomrule
    \end{tabular}
    }
    \raggedright
    \footnotesize
    $^1$Overall FPS includes both tracking and mapping iterations. \\
    $^2$Photo-SLAM uses feature point matching and thus achieves higher throughput.
    \label{tab:slam_comparison}
\vspace{-15pt}
\end{table}

\section{Profiling and Analysis}
\label{sec:motivation}

In this section, we first identify the underlying causes of the suboptimal speed performance of the 3DGS-SLAM pipeline, and then analyze the hardware inefficiencies encountered when deploying 3DGS-SLAM on GPUs, highlighting the key observations with \textbf{multi-level redundancies} that motivate our design. 

Our profiling contains three datasets: TUM-RGBD ($480\times640$ resolution)~\cite{TUM}, Replica ($680\times1200$ resolution)~\cite{Replica}, and ScanNet ($968\times1296$ resolution)~\cite{Scannet}, each providing color and depth images  of diverse indoor scenes, and is widely adopted for evaluating SLAM systems. 
All experiments use the SOTA ONX edge GPU~\cite{ONX}.

\begin{figure}[t]
    \centering
    \includegraphics[width=1.0\linewidth]{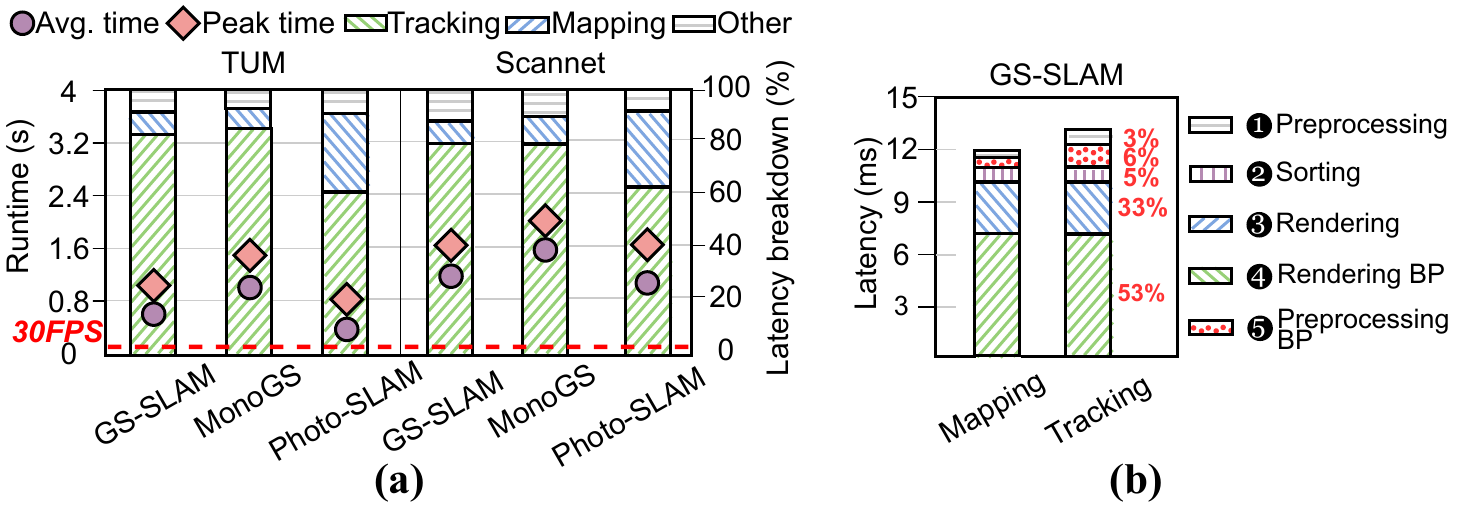}
    \vspace{-22pt}
    \caption{Latency breakdown of three SOTA 3DGS-SLAMs (including GS-SLAM~\cite{GSSLAM}, MonoGS~\cite{MONOGS}, and Photo-SLAM~\cite{photoslam}) across two datasets (including TUM~\cite{TUM} and Scannet~\cite{Scannet}) on ONX edge GPU~\cite{ONX}: (a) Latency breakdown of different stages in the SLAM pipeline and (b) Latency breakdown for tracking and mapping in GS-SLAM during a typical single iteration.}
    \Description{This figure shows the latency breakdown of state-of-the-art 3DGS-SLAM on the ONX edge GPU. Subfigure (a) provides the latency breakdown of different processes in the SLAM pipeline, while subfigure (b) details the latency of the tracking and mapping stages.}
    \label{fig:SLAM_profiling1}
\vspace{-15pt}
\end{figure}


\underline{\textbf{Pipeline-level Profiling.}}
We conduct a comprehensive analysis of three SOTA keyframe-based 3DGS-SLAM algorithms discussed in Sec.~\ref{sec:3DGS-SLAM_bg}: GS-SLAM~\cite{GSSLAM}, MonoGS~\cite{MONOGS} and PhotoSLAM~\cite{photoslam}. Each algorithm applies its own set of optimizations, aside from the original tracking and mapping procedures. To enable a fair comparison, we divide the overall pipeline into three main stages: tracking, mapping, and others. Fig.~\ref{fig:SLAM_profiling1}(a) shows the proportion of total runtime allocated to each stage. Based on the MonoGS algorithm, we further break down the runtime of the tracking and mapping stages under the TUM fr1/desk scene, as shown in Fig.~\ref{fig:SLAM_profiling1}(b).

\textbf{Observation 1: Tracking and Mapping are the primary bottlenecks in 3DGS-SLAMs.} 
Across the three algorithms, tracking and mapping stages together account for over 80\% of the total runtime across various scenarios (see Fig.~\ref{fig:SLAM_profiling1}(a)). Our profiling shows that tracking and mapping have similar per-frame latencies  (Fig.~\ref{fig:SLAM_profiling1}(b)), as both are configured with 50 iterations per frame. However, tracking runs on every frame, while mapping is only invoked on keyframes. This leads to a higher overall tracking time, even though their per-frame latency costs are comparable. 
To achieve real-time performance at 30 FPS, the overall system must be accelerated by more than 20$\times$.  
Therefore, achieving real-time performance on edge devices requires acceleration of both the tracking and mapping stages.

\textbf{Observation 2: Rendering and Rendering BP dominate the cost of both tracking and mapping stages.}  
As shown in Fig.~\ref{fig:SLAM_profiling1}(b), the time breakdowns for tracking and mapping stages exhibit similar patterns.  
In particular, {Step}~\ding{184} Rendering and {Step}~\ding{185} Rendering BP are the dominant components, accounting for over 80\% of overall runtime in both tracking and mappin stages.  

\begin{figure}[b]
    \centering
    \vspace{-10pt}
    \begin{minipage}{0.5\linewidth}
        \centering
        \includegraphics[width=\linewidth]{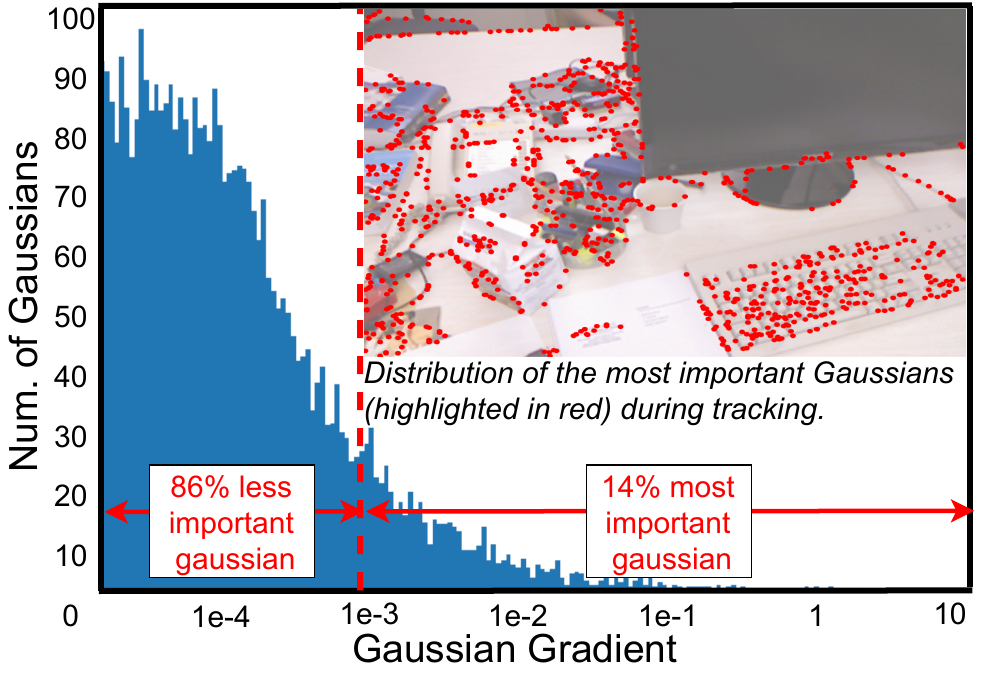}
        \vspace{-20pt}
        \caption{Gaussian gradient distribution during tracking.}
        \label{fig:Gaussian_redundant}
    \end{minipage}
    \hfill
    \begin{minipage}{0.44\linewidth}
        \centering
        \includegraphics[width=\linewidth]{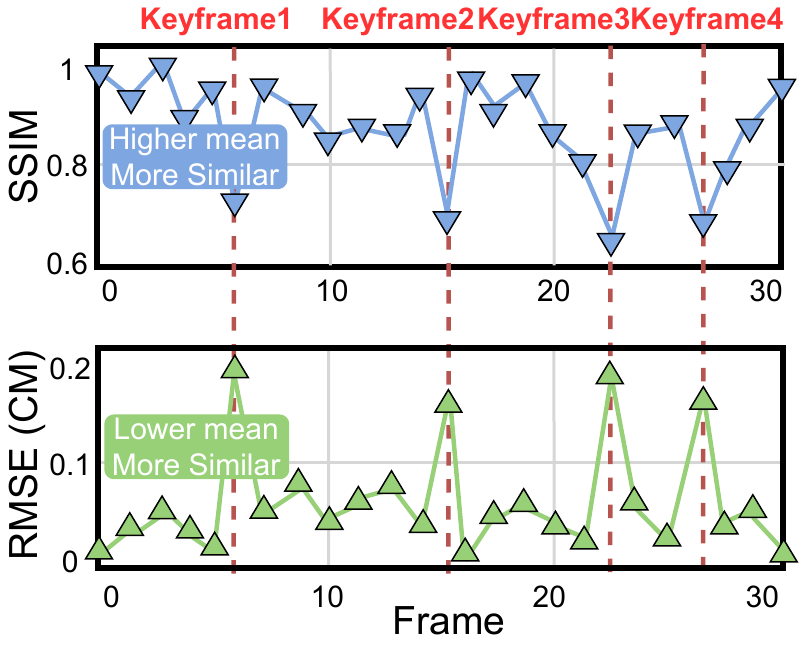}
        \vspace{-20pt}
        \caption{Similarity analysis in consecutive frames.}
        \label{fig:Pixel_redundant}
    \end{minipage}
    \label{fig:redundant}
\end{figure}

\underline{\textbf{Step-level Profiling.}}
We conduct step-level profiling of the tracking process in MonoGS~\cite{MONOGS}, a representative 3DGS-SLAM algorithm, on the TUM dataset~\cite{TUM}.  
MonoGS optimizes the camera pose/scene representation model using 3D Gaussian gradients, a method shared by most SOTA algorithms. 
To quantitatively demonstrate the contribution of each Gaussian to the pose optimization during tracking, we compute the gradient of each Gaussian and present the profiling results in Fig.~\ref{fig:Gaussian_redundant}.



\textbf{Observation 3: A substantial portion of Gaussians contributes negligibly to camera pose optimization.}  
We observe a highly skewed Gaussian gradient distribution as shown in Fig.~\ref{fig:Gaussian_redundant}, where a small fraction of Gaussians contribute significantly to the camera pose optimization.  
Specifically, only the top 14\% of Gaussians contribute the majority of the gradient magnitude, while the remaining 86\% of Gaussians exhibit negligible impact.  
Moreover, these important Gaussians are spatially clustered around the object contours and textured regions that are critical for pose estimation.
This observation indicates that a large number of Gaussians are less important during tracking, resulting in unnecessary overhead. 

\textbf{Observation 4: Rendering BP suffers from high latency due to massive atomic add operations and imbalanced pipeline.}  
We observe that the Rendering Backpropagation stage exhibits significantly higher latency compared to the forward rendering step during the tracking process.  
The primary cause lies in a large number of gradients concurrently updating Gaussian parameters at the same address, causing severe memory conflicts. To ensure correctness, atomic add serializes these updates, but the introduced stalls cause significant overhead and make this step a critical performance bottleneck in the tracking pipeline.
In addition, the pipeline suffers from imbalance, where the alpha gradient computation dominates runtime compared to other components during backpropagation. Although some of the parameters used in this computation are already available from the forward pass, the current design overlooks reuse opportunities, further prolonging latency.

\textbf{Can we directly prune all less important Gaussians within a single tracking frame?}  
Given the existence of a large number of Gaussians with small gradients during the tracking process, a natural question arises: can we directly prune all these less important Gaussians within a single tracking frame to accelerate computation?  
However, this is non-trivial. The gradient of a Gaussian reflects its contribution only in the current iteration, and its importance may change in subsequent iterations or under different camera views.  
Directly removing Gaussians based on their instantaneous gradient values may lead to suboptimal camera pose optimization or even tracking failure.  
Therefore, it is crucial to design a more careful and progressive pruning strategy that can dynamically remove unimportant Gaussians while preserving tracking accuracy.

\underline{\textbf{Frame-level Profiling.}}
We quantify inter-frame changes to better understand the SOTA keyframe-based mapping strategy and to expose potential optimization opportunities. Specifically, we choose MonoGS on the TUM-RGBD dataset and measure two metrics: (1) Root Mean Square Error (RMSE)~\cite{RMSE} for pixel-wise difference, where lower values indicate higher similarity in brightness; and (2) Structural Similarity Index Measure (SSIM)~\cite{SSIM} for structural similarity, where higher values indicate greater structural similarity.

\textbf{Observation 5: Redundant Computation in Non-keyframes.}  
As shown in Fig.~\ref{fig:Pixel_redundant}, consecutive frames exhibit high similarity, especially between non-keyframes.  
This observation suggests that treating all frames equally in terms of resolution and computation introduces unnecessary overhead in SLAM systems.  
In particular, keyframes are essential for maintaining accurate scene reconstruction and camera pose estimation, while non-keyframes mainly assist tracking with highly redundant content.  
Moreover, non-keyframes that are closer to keyframes tend to have higher similarity due to smaller camera motion and less scene variation, allowing for more aggressive resolution reduction.  
In contrast, as the distance from keyframes increases, the accumulated pose drift and scene changes become larger, requiring a gradual increase in resolution to preserve tracking accuracy and robustness.  
These observations motivate an adaptive computation strategy that dynamically adjusts the resolution of each frame based on its distance to the nearest keyframe, aiming to balance efficiency and accuracy.

\underline{\textbf{Iteration-level Profiling.}}
Workload imbalance across pixels within a frame has been noted in prior works, which can lead to low hardware utilization under a fixed pixel-to-hardware mapping~\cite{GauSPU,lin2025metasapiens}. Existing solutions rely on on-the-fly per-frame analysis to enable dynamic mapping, but this introduces a dilemma between the overhead of analysis and the workload balance. Unlike inference, in SLAM each frame executes Step~\ding{182} through Step~\ding{186} multiple iterations (typically {15-to-100} iterations per frame). We profile iteration-level workload distributions across pixels to highlight a unique opportunity in SLAM: reducing optimization overhead by reusing the workload distribution information across iterations.

\begin{figure}[t]
\vspace{-10pt}
    \centering
    \includegraphics[width=1.0\linewidth]{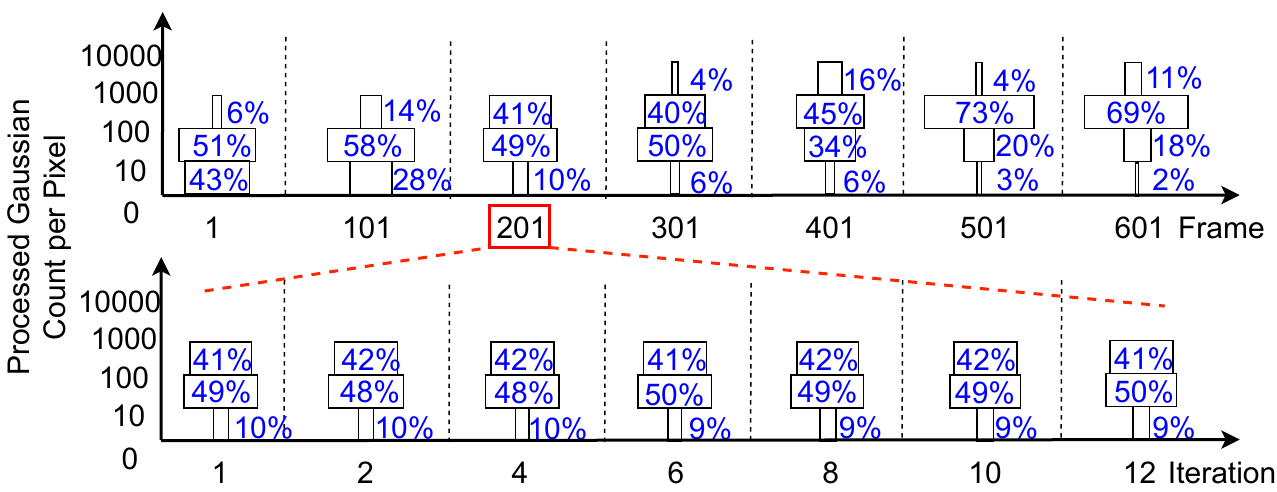}
    \vspace{-20pt}
    \caption{Workload distribution among pixels. The top figure shows the evolution of workload distribution across different frames. The bottom figure presents the workload distribution changes within one representative frame (Frame 201).}
    \label{fig:intersection}
\vspace{-10pt}
\end{figure}

\textbf{Observation 6: Similar Workload Distribution across Iterations.}
Fig.~\ref{fig:intersection} shows workload distributions measured as the number of Gaussians processed per pixel. Although workload distributions vary across frames, the distributions of consecutive iterations within a frame are highly similar. This is because tracking only updates the camera pose without modifying the Gaussians, causing each pixel's workload to change gradually across iterations. This similarity lets us reuse workload information to gradually adjust the distribution and converge to an optimal mapping after several iterations. In addition, we reuse the results of Step~\ding{182}-2 \textit{Tile Intersection} and Step~\ding{183} \textit{Sorting} to cut down computation overhead.

\section{{\name}: Algorithm}
\label{sec:alg}

\subsection{Adaptive Gaussian Pruning}
\label{sec:alg_pruning}
\textbf{Motivation.} Building on Observations 3 and 7, we design a \emph{progressive pruning strategy} that incrementally removes less important Gaussians across iterations. This approach maintains stable tracking accuracy while significantly reducing redundant computations.

\textbf{Algorithm.} The adaptive pruning algorithm uses gradient-based importance evaluation to retain Gaussians most critical to tracking. For each Gaussian, we consider two key factors that influence its contribution to the loss function: the mean position and the covariance scale. The corresponding gradients of the loss function to position and covariance are indicated by $d \mathcal{L} / d \boldsymbol{\mu}$ and $d \mathcal{L} / d {\Sigma}$.

The loss function \( \mathcal{L} \) is defined as the weighted sum of photometric and geometric residuals between the rendered image (produced by the Gaussian model) and the ground truth image. It is given by:
\begin{align}
\mathcal{L} = \lambda_{\text{pho}} E_{\text{pho}} + (1 - \lambda_{\text{pho}}) E_{\text{geo}},
\label{eq:loss}
\end{align}
where \( E_{\text{pho}} \) is the photometric residual, which measures the difference in pixel colors between the rendered and ground truth images, and \( E_{\text{geo}} \) is the geometric residual, which measures the difference in depth values. The weight \( \lambda_{\text{pho}} \) controls the relative importance of these two terms. This loss function is the optimization objective already required by 3DGS-SLAM during the tracking process, thus our pruning algorithm leverages the existing gradient computation without introducing additional loss calculation overhead.

To comprehensively evaluate the contribution of each Gaussian, we calculate the \( \ell_2 \)-norm of the gradients with respect to both the 3D mean and the covariance matrix ${\Sigma}$ to assess their importance.

Next, we combine these two norms in a weighted manner to quantify each Gaussian's overall impact on the loss. We define the importance score of each Gaussian as:
\begin{align}
\text{Score}_{\text{gaussian}} = \left\lVert \frac{d \mathcal{L}}{d \boldsymbol{\mu}} \right\rVert + \lambda \times \left\lVert \frac{d \mathcal{L}}{d \Sigma} \right\rVert,
\label{eq:importance_score}
\end{align}
where \( \lambda \) is used to balance influence of the position and scale.

Rather than pruning Gaussians with low importance scores in every iteration, we adopt a mask-prune strategy. Over \( K \) iterations, we mask Gaussians with low importance scores, excluding them from participating in the rendering process. In the \( (K{+}1)^\text{th} \) iteration, these Gaussians are permanently removed. The pruning interval \( K \) is dynamically adjusted, starting from an initial value \( K_0 \). After \( K_0 \) iterations, we calculate the change ratio of tile-Gaussian intersections. If this ratio exceeds 5\%, the next pruning interval is reduced to \( K_0/2 \); otherwise, it is increased to \( 2 \times K_0 \).

As shown in Sec.~\ref{sec:motivation}, the intersection relationships between tiles and Gaussians remain relatively stable across adjacent iterations, allowing us to reduce the time spent on sorting and preprocessing during the \( K \) iterations, which is critical for the subsequent structural design. We adopt the mask-prune strategy over direct pruning to preserve Gaussians for computing the tile-Gaussian change ratio.

\begin{figure*}[t]
    \centering
    \includegraphics[width=1.0\linewidth]{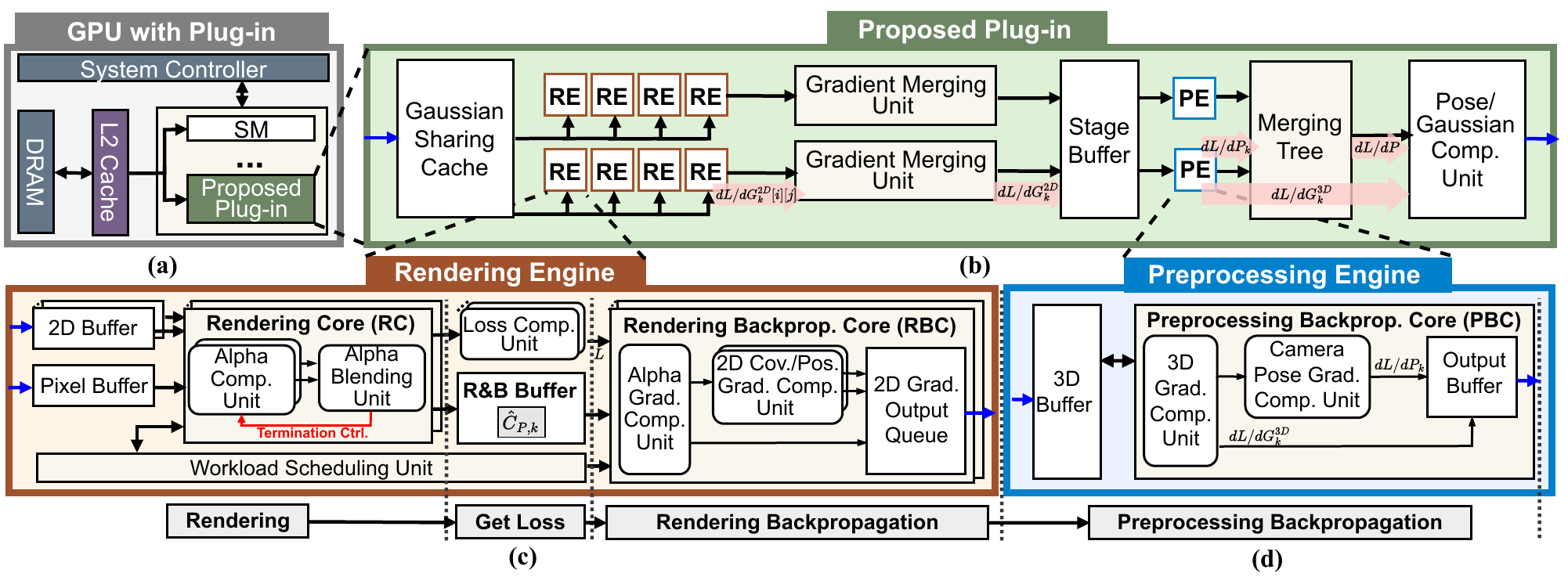}
    \vspace{-2.6em}
    \caption{Overall design of our acceleration system: (a) an illustration of the integration of {\name} plug-in with the GPU. {\name} plug-in shares the L2 cache with the GPU, (b) the overview architecture of {\name}; (c) the block diagram of the Rendering Engine (RE), responsible for \textbf{Step}~\ding{184} Rendering, loss, and \textbf{Step}~\ding{185} Rendering BP; and (d) the block diagram of the Preprocessing Engine (PE), responsible for \textbf{Step}~\ding{186} Preprocessing BP.}
    \label{fig:hardware_overall_arch}
    \vspace{-10pt}
\end{figure*}

Compared to existing Gaussian pruning methods, such as LightGaussian~\cite{LightGaussian} and MaskGaussian~\cite{Maskgaussian}, which rely on additional metrics or heuristics to estimate the importance of each Gaussian, our approach directly exploits the gradient information generated during the 3DGS-SLAM optimization process. Since the computation of gradients with respect to Gaussian parameters is already an integral part of camera pose optimization, our method introduces no extra computational overhead for importance evaluation, enabling efficient and lightweight pruning without impacting performance.

\subsection{Dynamic Downsampling}
\label{sec:alg_downsampling}
\textbf{Motivation.} Observation 5 indicates that processing all frames at high resolution leads to significant redundant computation. To address this, we design a \emph{dynamic downsampling} technique that adaptively adjusts each frame's resolution based on its  importance, reducing  overhead without compromising tracking accuracy.

\textbf{Algorithm.} Instead of processing all frames at a fixed high resolution as in vanilla 3DGS-SLAM, our dynamic downsampling framework adaptively lowers the resolution for less critical frames, improving efficiency without compromising performance.

Our framework processes \textbf{keyframes} at the original full resolution \( R_0 \), ensuring that crucial mapping and localization information is captured with high precision. For \textbf{non-keyframes}, we reduce the computational burden by downsampling them. Specifically, when a non-keyframe follows a keyframe, its resolution is reduced to $(1/16)R_0$. If subsequent frames continue to be non-keyframes, their resolution is incrementally increased by a scaling factor \( m > 1 \), up to a maximum of $(1/4)R_0$. This gradual increase continues for each consecutive non-keyframe until a new keyframe is selected, at which point the resolution is reset to \( R_0 \).

Mathematically, the resolution \( R_n \) of frame \( n \) is determined as:

\textbf{For keyframes:} $R_n = R_0$

\textbf{For Non-keyframes:} $R_n = \min\left( (1/16) R_0 \times m^{(n - k - 1)}, \, (1/4) R_0 \right)$
where \( k \) denotes the index of the most recent keyframe.

This adaptive resolution strategy enables us to optimize computational resource allocation by reducing the processing load during periods when high-resolution data is not critical. On non-keyframes, we adopt a progressive downsampling approach instead of abrupt resolution drops, thereby ensuring a smooth transition in visual quality.
This gradual adjustment helps prevent trajectory drift and ensures stable SLAM optimization, whereas abrupt resolution changes often lead to accuracy degradation. We always process keyframes at full resolution to capture essential scene information. For non-keyframes, the resolution is progressively reduced based on their temporal distance to the nearest keyframe: the closer a frame is to a keyframe, the more similar its content and the less new information it provides, thus requiring lower resolution. Experimental results show that with our method, both ATE and PSNR remain within a 10\% variance across all sequences, demonstrating robustness and balanced trade-off between efficiency and accuracy.

\section{{\name}: Architecture}
\label{sec:arch}


In this section, we propose the {\name} architecture, which is a plug-in that can be integrated into GPUs to accelerate 3DGS-SLAM workloads. As shown in Fig. ~\ref{fig:hardware_overall_arch}(a), the integration system has two components: (1) a \textbf{{\name} plug-in} that leverages co-design techniques for both tracking and mapping stages (Fig. ~\ref{fig:hardware_overall_arch}(b)), and (2) the original \textbf{GPU} that accelerates {Step}~\ding{182} Preprocessing and {Step}~\ding{183} Sorting, enhancing end-to-end performance of SLAM applications. In the following, we first introduce the {\name} architecture overview (Sec.~\ref{sec:arch_1}), then the designs of three key units (Sec.~\ref{sec:rendering_enging}, Sec.~\ref{sec:arch_GMU}, and Sec.~\ref{sec:arch_preprocessing}), and finally the system integration with GPUs (Sec.~\ref{sec:programming_model}).

\subsection{{\name} Architecture Overview}
\label{sec:arch_1}
\textbf{Architecture Overview.} Fig.~\ref{fig:hardware_overall_arch}(b) illustrates the overall architecture of {\name}, which comprises four main hardware modules: Rendering Engines (\textbf{RE}) (for {Step}~\ding{184} Rendering and pixel-level 2D Gaussian gradients in {Step}~\ding{185} Rendering BP), Gradient Merging Units (\textbf{GMU}) (for 2D Gaussian gradients in {Step}~\ding{185}), Preprocessing Engines (\textbf{PE}) (for 3D Gaussian updates in {Step}~\ding{186} Preprocessing BP), and a Pose Computing Unit (for camera pose update in {Step}~\ding{186}). We further insert a Gaussian Sharing Cache for sharing 2D/3D Gaussian attributes with GPU Streaming Multiprocessors (SMs), a Stage Buffer for caching 2D Gaussian gradients between GMUs and PEs, and a Merging Tree for accumulating camera pose gradients.

\textbf{Overall 3DGS-SLAM to RTGS Architecture Mapping.} 
We introduce the overall 3DGS-SLAM algorithm to RTGS architecture mapping with an emphasis on input/output data flow across modules. Similar to prior GPU implementations and 3DGS accelerators, RTGS supports parallel computation over pixel grids. However, with the optimizations, RTGS only requires parallel computation over 16$\times$16 pixels, equivalent to one pixel grid or one tile in most prior designs~\cite{GauSPU,distwar}, to achieve real-time performance. For this, we follow this convention for defining a tile and refer to the smallest parallel compute unit in RTGS, consisting of 4$\times$4 pixels, as a subtile. 

During the {Step}~\ding{184} Rendering and {Step}~\ding{185} Rendering BP stages, each RE processes a subtile of 16 pixels and generates the corresponding pixel-level 2D Gaussian gradients ($dL/dG^{2D}_k[i][j]$) within this subtile. To mitigate the workload imbalance across subtiles, the REs operate in a streaming fashion, where subtile-level workloads are dispatched once REs are free and resulting in asynchronous workload dispatching. 
After REs, we insert GMUs to merge gradients for the same 2D Gaussian within each subtile and only pass the 2D Gaussians ($dL/dG^{2D}_k[m]$) to the following Stage Buffer. The merged 2D Gaussian gradients of each subtile are stored and accumulated in the Stage Buffer, preparing the Gaussian-level 2D gradients ($dL/dG^{2D}_k$) for {Step}~\ding{186} Preprocessing BP.
PEs receive 2D Gaussian gradients from the Stage Buffer and then compute the transformation from 2D gradients ($dL/d^{2D}_k$) to 3D gradients ($dL/d^{3D}_k$) for each Gaussian during {Step}~\ding{186} Preprocessing BP. During tracking stages, PEs further take 3D gradients and compute the pose gradients from each 3D Gaussian ($dL/dP_k$). These pose gradients ($dL/dP_k$) are merged using the Merging Tree ($dL/dP$) and sent to the Pose Computing Unit to optimize the camera pose.

\subsection{Rendering Engine}
\label{sec:rendering_enging}

\textbf{Motivation.} The Rendering Engine (RE) is responsible for handling {Step}~\ding{184} Rendering and {Step}~\ding{185} Rendering BP, which characterizes the primary performance bottlenecks in the 3DGS-SLAM, as illustrated in Observation 2. Through algorithm analysis, we observe that alpha gradient computing dominates {Step}~\ding{185} Rendering BP, which exhibits longer latency than {Step}~\ding{184} Rendering. This overhead arises from recomputing the alpha and transmittance value using time-consuming division operation (Eq. \ref{eq:back_2}), which is calculated in (Eq. \ref{eq:color}) using multiplications but discarded. Moreover, the imbalance of workload between different pixels further causes suboptimal hardware utilization, thus increasing the overall execution time. Consequently, balancing the execution pipeline and mitigating workload imbalance are our main design considerations in RE.

\textbf{RE Design Overview.} Motivated by the aforementioned redundancies, each RE design involves 8 Rendering Cores (RCs) and Rendering Backpropagation Cores (RBCs) with balanced resource allocation, and an Rendering \& Backpropagation (R\&B) Buffer for cross-stage data reuse. A Workload Scheduling Unit (WSU) is integrated into the RE to address workload imbalance within a subtile.

\textbf{Rendering Core (RC).} As shown in Fig.~\ref{fig:hardware_overall_arch}(c), a Rendering Core (RC) consists of an alpha computing unit and an alpha blending unit to execute fragment-level operations with an RE (see Eq.~\ref{eq:alpha} and Eq.~\ref{eq:color}). The pixel buffer stores the pixels, whereas the 2D FIFO retains the 2D Gaussians for the next few rounds. Given these inputs, the alpha computing units in RCs compute alpha values of each fragment. These alpha values are sent to the alpha blending units to compute the pixel color determined by all intersected Gaussians, with an early termination control signal. When the transmittance value $T$ of a fragment reaches a predefined threshold, the early termination control signal terminates the iteration rendering of the pixel. 

Motivated by the significant latency gap between {Step}~\ding{184}-1 Alpha Computing and {Step}~\ding{184}-2 Alpha Blending, 12 and 3 cycles, respectively, and the data dependency inherent in the execution, we mitigate pipeline imbalance through resource reallocation. Specifically, each pixel is assigned one dedicated alpha computing unit, while a single alpha blending unit is shared between two pixels. This approach reduces hardware cost, while the extra increase (3 cycles) in execution can be hidden by pipeline balancing. 

\textbf{Rendering Backpropagation Core (RBC).} After getting pixel loss, RBC computes the fragment-level 2D color, covariance and position gradients, as shown in Fig.~\ref{fig:hardware_overall_arch}(c). The alpha gradient computing unit computes from color loss to alpha gradient $dL/d\alpha$. After obtaining $dL/d\alpha$, the pipeline continues with covariance/position gradient computing $dL/d\Sigma$ and $dL/dpos$. The $dL/dC_P$ computed by the loss computing unit, with $dL/d\Sigma$ and $dL/dpos$ are stored in the 2D gradient output queue, waiting for gradient merging.

\textbf{R\&B Buffer.} Due to the time-consuming alpha recomputation, the alpha gradient computing  proposed in baselines requires 20 clock cycles, significantly exceeding the 8 cycle latency of other stages in the {Step}~\ding{185} Rendering BP pipeline. However, as shown in Eq. \ref{eq:color} and \ref{eq:back}, the variable $\hat{\mathbf{C}}_{P,k} = T_{P,k} \alpha_{P,k} \mathbf{C}_k$ required for this computation is available from the forward. To exploit this opportunity, we introduce an R\&B Buffer to enable parameter reuse, as illustrated in Fig. \ref{fig:hardware_overall_arch}(c). With reuse enabled, the latency of the alpha gradient computating is reduced from 20 cycles to 4 cycles. 
 

\begin{figure}[t]
    \centering   \includegraphics[width=1.0\linewidth]{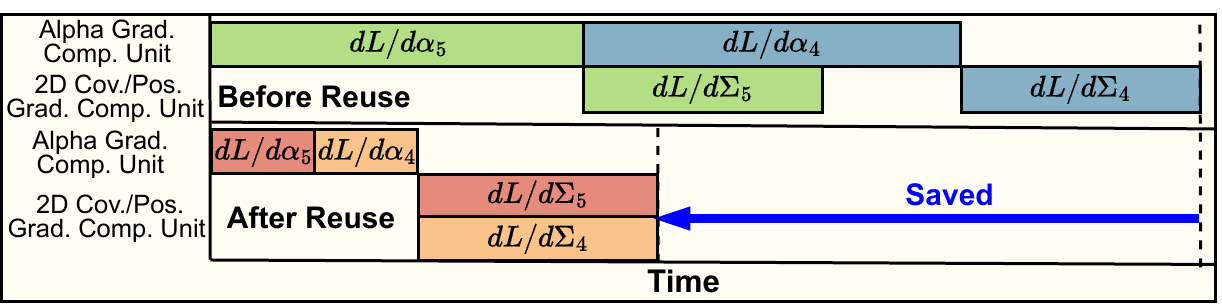}
    \vspace{-22pt}
    \caption{A timing diagram to illustrate the influence of parameter reuse on the pipeline of {Step}~\ding{185} Rendering BP.}
    \label{fig:reuse_mot}
\vspace{-20pt}
\end{figure}

The working mechanism of the R\&B Buffer leverages a double-buffered structure to enable concurrent read and write operations. Data are managed and transferred at the granularity of chunks, and each chunk typically contains four intermediate values $\hat{\mathbf{C}}_{P,k}$ per pixel. During {Step}~\ding{184} Rendering, the computed $\hat{\mathbf{C}}_{P,k}$ values are written back to the Gaussian Cache along with their corresponding pixel ID and chunk ID, where the chunk ID denotes the execution order. While the current chunk is consumed for alpha gradient computation, the next chunk is concurrently prefetched onto the R\&B buffer. This chunk-level preloading mechanism ensures that the data stored in the R\&B Buffer remains constant during execution. Furthermore, since the data loading latency from the Gaussian Cache to the R\&B Buffer is shorter than the compute latency, the dedicated Gaussian Cache and R\&B Buffer design together ensure a high-throughput execution pipeline without memory overhead.

Similar to RC, RBC incorporates resource reallocation to improve the pipeline, employing one shared alpha gradient computing unit along with dedicated units for 2D covariance and position gradients. Since computing $dL/d\Sigma$ and $dL/dpos$ takes 8 cycles while computing $dL/d\alpha$ is reduced to 4 cycles, we derive a balanced pipeline during {Step}~\ding{185} Rendering BP in~\name, as shown in Fig. \ref{fig:reuse_mot}.

\textbf{Workload Scheduling Unit.} To avoid REs dominating computation, we design the Workload Scheduling Unit (WSU), which combines intra-RE scheduling with inter-RE streaming to reduce imbalance and leverage inter-iteration similarity for scheduling.

\begin{figure}[t]
    \centering
    \includegraphics[width=1.0\linewidth]{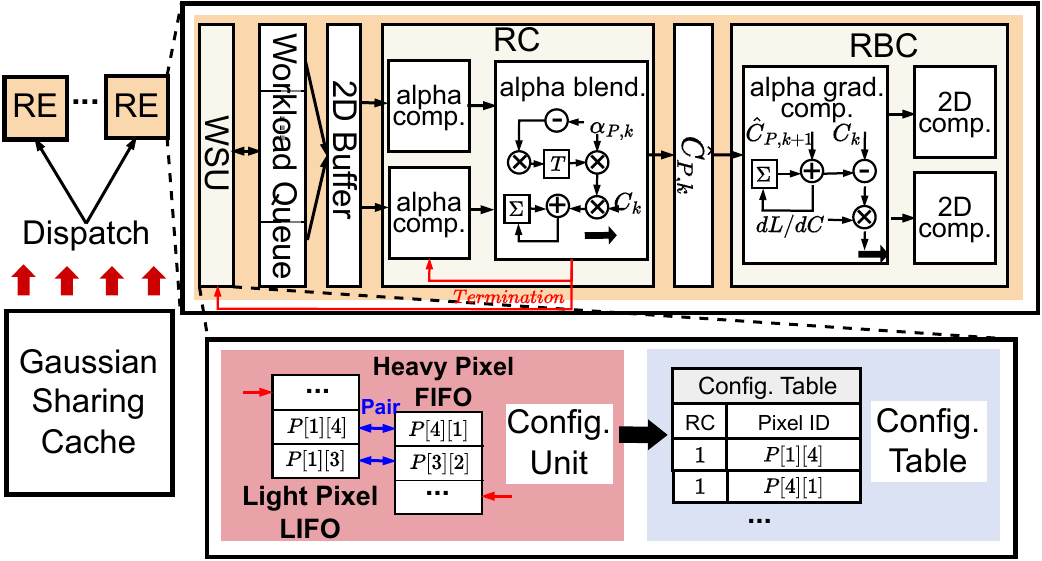}
    \vspace{-20pt}
    \caption{The block diagram of the RE with an illustration of workload scheduling realized by WSU during {Step}~\ding{184} Rendering and {Step}~\ding{185} Rendering BP.}
    \label{fig:wsu}
 \vspace{-10pt}
\end{figure}

As depicted in Fig.~\ref{fig:wsu}, on the intra-RE level, the current round of workload is dispatched and stored in the workload queue. The WSU retrieves configuration data from the previous iteration and assigns workloads to corresponding 2D buffers via a fully connected network. Initially, each buffer concurrently transmits two Gaussians, each intersected with a distinct pixel, to the RC. Once all Gaussians associated with a given pixel have been processed, the alpha blending unit issues a termination signal not only to alpha computing unit, but also to the WSU, indicating the completion of execution for that pixel. Subsequently, the RC continues by concurrently processing two Gaussians associated with the same uncompleted pixel, blending them in a sequential order. In this manner, the two pixels sharing the same 2D buffer will complete their computations simultaneously, thereby balancing utilization of computational resources. To mitigate excessive scheduling overhead, here we adopt a pairwise workload balancing strategy between pixels.

\begin{figure}[b]
\vspace{-10pt}
    \centering
    \includegraphics[width=1.0\linewidth]{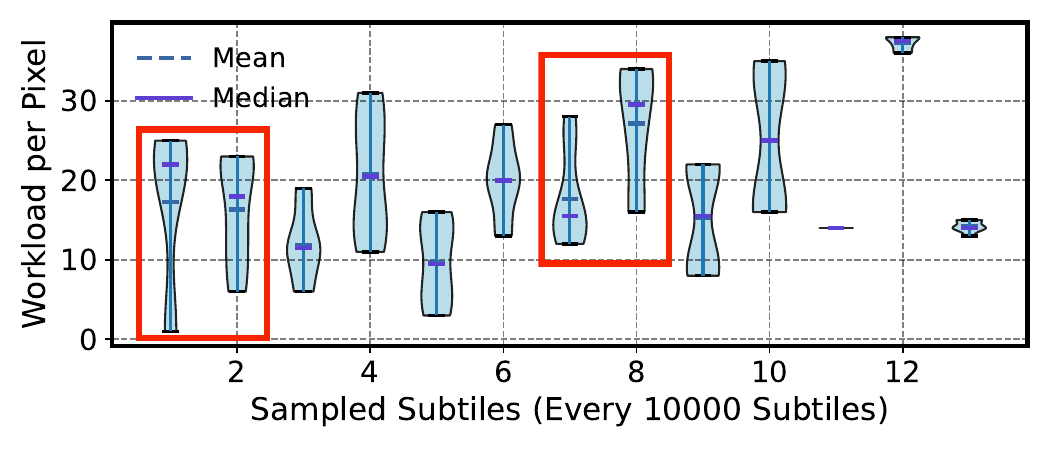}
    \vspace{-25pt}
    \caption{Illustration of the choice of pairwise scheduling.}
    \label{fig:box}
\end{figure}

However, simply switching between two adjacent pixels does not lead to significant performance improvement. To address this, we leverage inter-iteration similarity by utilizing the scheduling information from the previous iteration to guide the pixel pairing of the current iteration. Specifically, as shown in the violin plot of Fig. \ref{fig:box}, pixels with excessively high and low workloads tend to be symmetrically distributed within most subtiles. The asymmetric ones only account for 11\% of all the subtiles, thereby making the speedup achieved by pairwise scheduling approaches close to the ideal one. Based on this observation, we pair high-workload pixels with low-workload ones and assign each pair to the same 2D buffer.

Since the termination signal provides the completion order of each pixel, and no sorting and intersection are performed between iterations, this order remains largely consistent across iterations. In each iteration, we record the completion order of 8 light-workload pixels in a FIFO buffer. Once the FIFO is full, we begin recording the completion order of 8 heavy-workload pixels in a LIFO buffer. Both queues are populated in ascending order of completion time. Since their popping follows reverse access patterns, the pixel IDs output within the same cycle naturally constitute the heavy--light pairs. These pairs are stored in a configuration table, which is then used in the next iteration to guide the scheduling of the same subtile.

On the inter-RE level, we adopt a streaming approach where each RE asynchronously processes subtiles. Once an RE finishes, the next subtile is streamed in, enabling pipelined execution. While less effective than global tile-level scheduling in balancing workloads, streaming significantly alleviates critical scheduling issues caused by congestion and routing in large-scale interconnects.

Combining intra-RE scheduling with inter-RE streaming reduces workload imbalance by 33.06\% on average, with minimal overhead.

\begin{figure}[t]
    \centering
    \includegraphics[width=1.0\linewidth]{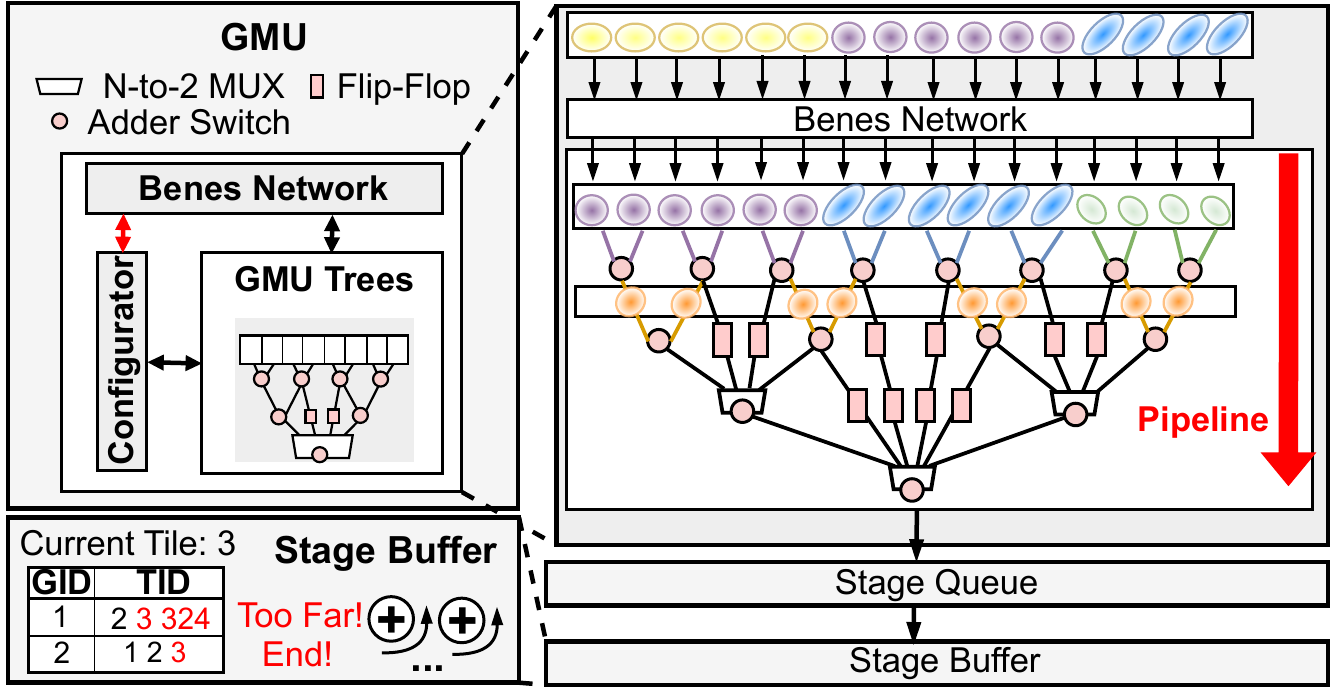}
    \vspace{-20pt}
    \caption{The block diagram of GMU.}
    \label{fig:gmu}
\vspace{-10pt}
\end{figure}

\subsection{Gradient Merging Unit} 
\label{sec:arch_GMU}

\textbf{Motivation.} Observation 4 highlights that the atomic gradient merging operations in GPU implementation cause severe memory conflicts. As Gaussians are more scattered in SLAM workloads, we need dedicated design for sparse gradient aggregation: from pixel-level $dL/dG^{2D}_k[i][j]$, to tile-level $dL/dG^{2D}_k[m]$, and finally Gaussian-level $dL/dG^{2D}_k$, to fully exploit aggregation opportunities.

\textbf{GMU Design Overview.} To fully leverage tile-level reduction opportunities for the scattered Gaussians to reduce memory conflicts, we introduce GMUs between REs and PEs to enable sparse gradient aggregation and updates for the same 2D Gaussian. The merged Gaussian gradients are further accumulated in the Stage Buffer for Gaussian-level aggregation.

\textbf{GMU.} 
Due to the situation that gradients from each RE may not be associated with the same Gaussian after scheduling optimization, we utilize a Benes Network to rearrange and cluster these gradients, as shown in Fig.  \ref{fig:gmu}. After rearranging, a unique reduction tree in the GMU is adopted from~\cite{sigma}, which introduces bypass links across multiple adder levels, thereby extending the traditional adder tree design to clustered aggregation. Starting from the second adder level, an N-to-2 multiplexer is placed before each adder. The routing of Gaussian clusters is controlled by the configurator, based on data collected from the previous stage.

For intra-tile (pixel-level to tile-level) aggregation, 16 REs are divided into four groups, each performs pipelined aggregation of gradients from 4 REs. Specifically, flip-flops are inserted along bypass paths to ensure synchronization, such that gradients from RE1 can be computed at the second level of the GMU tree simultaneously with gradients from RE2 at the first level. This merging scheme reduces latency and hardware overhead of the GMU.

For inter-tile (tile-level to Gaussian-level) aggregation, a stage queue is employed to temporarily buffer partially merged results, and the final gradient accumulation is completed in the stage buffer. For each Gaussian, once all its gradients have been aggregated or if its next occurrence is distant in the execution order, it is marked as evictable and can be replaced by a new Gaussian entry. Experiments show that the latency of gradient merging alone is reduced by 68.04\% on average compared to using atomic operations.


\subsection{Preprocessing Engine}
\label{sec:arch_preprocessing}
\textbf{Motivation.} {\name} accelerates for both tracking and mapping during Step~\ding{186} Preprocessing BP, where pose
gradients are further merged using the Merging Tree and
sent with 3D Gaussian gradients to the Pose/Gaussian Computing Unit for optimization.

\textbf{PE Design Overview.} We design each PE with a PBC to process both pose gradients $dL/dP_k$ and 3D Gaussian gradients $dL/dG^{3D}_k$ for tracking and mapping respectively, as shown in Fig.~\ref{fig:hardware_overall_arch}(d).

\textbf{Preprocessing Backpropagation Core (PBC).} Each PE includes one PBC to process Gaussian-level gradients fused by the GMU. The PBC receives gradient batches and stores either pose gradients $dL/dP_k$ or 3D Gaussian gradients $dL/dG^{3D}_k$ in the output buffer. In mapping, it stores the final 3D gradients; in tracking, pose gradients are further aggregated by the Merging Tree.

\begin{figure}[t]
    \centering
    \includegraphics[width=1.0\linewidth]{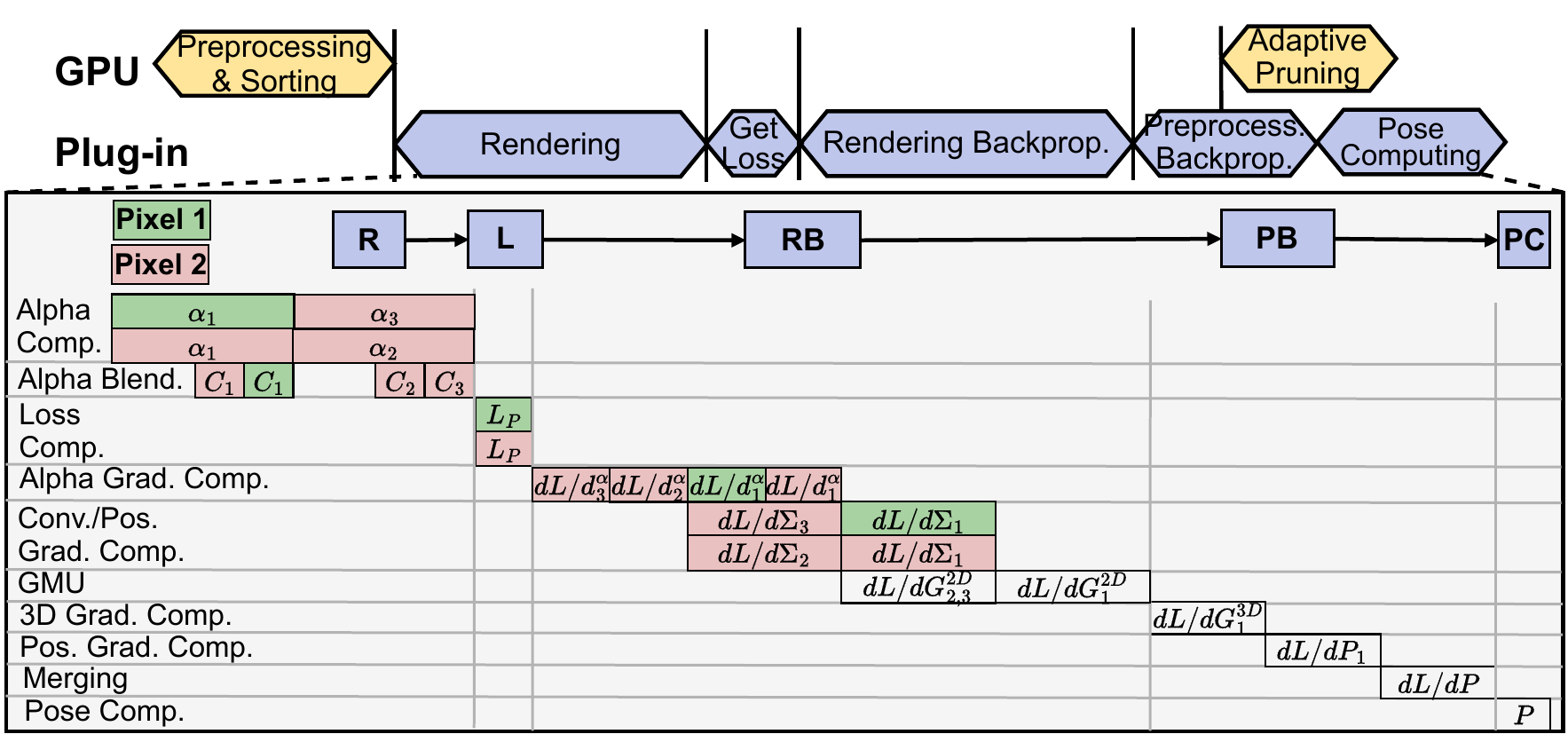}
    \vspace{-20pt}
    \caption{The GPU and plug-in integration pipeline.}
    \label{fig:pip}
 \vspace{-10pt}
\end{figure}
\subsection{Integration with GPUs}
\label{sec:programming_model}

\textbf{Workload Partitioning Between {\name} and GPU.}
We partition workloads to leverage both GPU and {\name}: the GPU handles {Step}~\ding{182} Preprocessing, {Step}~\ding{183} Sorting and Pruning, while {\name} accelerates {Step}~\ding{184} Rendering and {Step}~\ding{185}+~\ding{186} BP (Fig.~\ref{fig:pip}).



\textbf{Programming Model.} RTGS adopts a function-level interface (Listing~\ref{listing:api}) for coordinating with GPU SMs, inspired by GBU~\cite{GBU}. It exposes two core functions for execution and status checking, enabling modular acceleration in 3DGS-SLAM.

\texttt{RTGS\_execute(frame\_id, is\_keyframe)} is responsible for executing the processing of a single SLAM frame. It is called after GPU SMs complete preprocessing and Gaussian sorting. RTGS then performs rendering and backpropagation, computes gradients for each Gaussian, and writes them to shared memory, where SMs handle pruning (for non-keyframes). Synchronization between SMs and RTGS is managed via shared-memory flag buffers: RTGS first polls an \texttt{Input\_done} flag to detect when SMs finish preprocessing and sorting, then sets a \texttt{gradient\_ready} flag to notify SMs to start pruning. Once pruning completes, SMs write a \texttt{pruning\_done} flag, which RTGS polls before writing back results. For non-keyframes, RTGS writes the optimized pose to L2 cache; for keyframes, it skips pruning and pose update, and instead uses gradients to update Gaussian parameters for mapping.

\begin{figure}[t]  
\lstset{
    caption=C++ programming interface of {\name}.,
    language=C++,
    frame=single,
    label=listing:api,
    basicstyle=\footnotesize\ttfamily,
    columns=flexible,
    breaklines=true,
    tabsize=4,
    commentstyle=\color{gray}\ttfamily,
    moredelim=**[is][\color{olive}]{@olive@}{@endolive@}, 
}
\begin{lstlisting}
// Trigger RTGS execution for a single SLAM frame
void RTGS_execute(
    int frame_id,            // frame identifier
    bool is_keyframe,        // whether this is a keyframe
    const void* sorted_gaussians, // sorted Gaussians
    const void* image_data,       // input observation
    void* gradient_buffer,        // output gradient(to SMs)
    void* pose_buffer             // write-back camera pose
);
// Query current RTGS status for a given frame
int RTGS_check_status(
    int frame_id,            // frame identifier
    bool blocking            // Wait until RTGS is idle
);
\end{lstlisting}
\vspace{-1em}
\end{figure}

\texttt{RTGS\_check\_status(frame\_id)} reports the current execution status of RTGS (\texttt{IDLE}, \texttt{EXECUTING}, or \texttt{WAIT\_PRUNING}) and includes an optional blocking flag. This feature allows the host thread to wait for RTGS to complete the current frame before starting the next, ensuring proper coordination across frames without the need to rely on CUDA stream synchronization.



\section{Evaluation}
\label{sec:evaluation}

\subsection{Experiment Setup}
\label{sec:Exp}

\begin{table}[b]
\vspace{-10pt}
  \caption{Dataset setup for evaluation.}
  \vspace{-10pt}
  \centering
    \resizebox{0.9\linewidth}{!}
    {
      \begin{tabular}{c|c|c}
      
      \toprule[1pt]
       \textbf{Dataset} & \textbf{Scenes} & \textbf{Resolution} \\
      \midrule
      \midrule
      TUM-RGBD~\cite{TUM} & fr1/desk, fr2/xyz, fr3/office & 480 $\times$ 640\\
      \midrule
      \multirow{2}{*}{Replica~\cite{Replica}} &  Rm0, Rm1, Rm2, & \multirow{2}{*}{680 $\times$ 1200}\\
       & Off0, Off1, Off2, Off3 & \\
      \midrule
      \multirow{2}{*}{ScanNet~\cite{Scannet}} & scene0000, scene0059,  scene0106 & \multirow{2}{*}{968 $\times$ 1296}\\
       & scene0269, scene0181, scene0207 & \\
      \midrule
       ScanNet++~\cite{Scannet++} & s1, s2 & 1160 $\times$ 1752\\
      \bottomrule[1pt]
      \end{tabular}
      }
    \label{tab:dataset}
\end{table}


\textbf{Datasets.}
We evaluate the performance of our proposed {\name} on four commonly used visual SLAM datasets: TUM-RGBD~\cite{TUM}, Replica~\cite{Replica}, the ScanNet~\cite{Scannet}, and ScanNet++\cite{Scannet++}. Tab.~\ref{tab:dataset} summarizes their scenes and frame resolutions.

\textbf{{\name} Algorithm Setup.}
Our proposed {\name} algorithm techniques are general and can work as a plug-and-play extension to existing 3DGS-SLAM algorithms. To clarify, we denote the existing algorithms as \textit{base} algorithms. We chose \underline{three base 3DGS-SLAMs} \underline{using keyframe mapping}: GS-SLAM~\cite{GSSLAM}, MonoGS~\cite{MONOGS}, and Photo-SLAM~\cite{photoslam}. For Photo-SLAM with a traditional geometry-based tracking backpropagation, we only apply our techniques to its rendering and mapping backpropagation. Each base algorithm adopts a distinct keyframe selection strategy: GS-SLAM selects keyframes based on scene changes (e.g., pose distance), MonoGS uses fixed intervals between frames, and Photo-SLAM relies on photometric changes. We retain each algorithm's original keyframe policy. To ensure consistency, we use the fixed hyperparameters for our method: an adaptive pruning threshold $\lambda = 0.8$, an initial pruning interval $K_0 = 5$, and a downsampling scaling factor $m = 2$.

In order to make a fair comparison with the SOTA GauSPU~\cite{GauSPU}, we build {\name} algorithm on top of \underline{on base 3DGS-SLAM without} \underline{keyframe mapping}, i.e., SplaTAM~\cite{splatam}, as well. Specifically, we apply our techniques to the tracking iterations of each frame.

\textbf{Algorithm Baselines.}
The \underline{four base 3DGS-SLAM algorithms} serve as one set of algorithm baselines. In addition, as {\name} algorithm unifies Gaussian pruning and image pixel downsampling into one SLAM framework, we further benchmark over \underline{one Gaussian} \underline{pruning technique}, i.e., Taming 3DGS~\cite{Taming}, and  \underline{one sparse sampling} \underline{technique}, i.e., GauSPU~\cite{GauSPU}, to validate its superiority. Taming 3DGS provides open-source code\footnote{https://github.com/humansensinglab/taming-3dgs}, allowing evaluation across various datasets and base algorithms. Since GauSPU does not release its code, we compare using the same dataset and base SLAM algorithm it employs, i.e., Replica and SplaTAM~\cite{splatam}, respectively.

\begin{table}[t]
\centering
\renewcommand{\arraystretch}{1.2} 
\setlength{\tabcolsep}{5pt}
\caption{{\name} architecture configurations.}
\vspace{-8pt}
\resizebox{\linewidth}{!}{
\begin{tabular}{|>{\centering\arraybackslash}m{3cm}|>{\centering\arraybackslash}m{1.7cm}|>{\centering\arraybackslash}m{2.3cm}|>{\centering\arraybackslash}m{1.7cm}|}
\hline
Technology Node & 28nm & Operating Freq. & 500 MHz \\ \hline
Power & 8.11W  & Area & 28.41mm\textsuperscript{2} \\
\hline \hline
\multicolumn{4}{|c|}{\textbf{Computation Resources}} \\ \hline
\multirow{2}{=}{\centering RE $\times$ 16:\\ 8 RCs \& RBCs per RE } & \multirow{2}{=}{\centering WSU $\times$ 16} & \multirow{2}{=}{\centering PE $\times$ 16: \\ 1 PBC per PE } & \multirow{2}{=}{\centering GMU $\times$ 4} \\ 
 &  &  &  \\\hline
\multicolumn{4}{|c|}{\textbf{Memory Allocation}}\\ \hline
Gaussian Cache & 80KB & Pixel Buffer & 24KB \\\hline 2D Buffer & 20KB  & R\&B Buffer & 16KB \\ \hline
Stage Buffer & 16KB & 3D Buffer & 10KB \\ \hline
Output Buffer & 15KB & WSU Buffer & 16KB \\ \hline
SRAM & 197KB & L2 Cache & 2MB \\ \hline
\end{tabular}
}
\label{tab:specification}
\vspace{-10pt}
\end{table}

To evaluate the quality-performance trade-off, we include two more precise baselines: LightGaussian~\cite{LightGaussian} and FlashGS~\cite{FlashGS}. These methods retain more Gaussians by using multiple metrics, such as PSNR and image saliency maps, to guide importance evaluation. For fairness, we adopt their original experimental settings and apply a uniform 50\% pruning ratio across all methods.


\textbf{{\name} Hardware Setup.} 
Tab.~\ref{tab:specification} shows the {\name} hardware module configurations.
The shape of one tile is set as 16$\times$16 pixels, which is further divided into 16 sub-tiles of 4$\times$4 pixels each. Our {\name} adopts 16 REs and 16 PEs, with each RE executing one subtile and each PE processing 16 Gaussians in parallel. 
The {\name} area and power are summarized in Tab.~\ref{tab:device_specs}. We implemented the proposed {\name} architecture hardware in Verilog and synthesized based on 28nm technology using the Synopsys Design Compiler~\cite{DC} and memory compiler from the vendor. The area data are from DC and memory compiler. The typical power consumption is as reported by Synopsys PrimePower~\cite{Primepower} based on the generated gate-level netlist and Verilog simulation for the target datasets.

\begin{table}[t]
    \centering
    \renewcommand{\arraystretch}{1}
    \caption{Comparison of device specifications.}
    \vspace{-8pt}
    \resizebox{\columnwidth}{!}{ 
    \begin{tabular}{c c c c c c} 
        \toprule
        \textbf{Device} & \textbf{Technology} & \textbf{SRAM} & \makecell{\textbf{Number} \\ \textbf{of Cores}} & \makecell{\textbf{Area} \\ \textbf{[$mm^2$]}} & \makecell{\textbf{Power} \\ \textbf{$[W]$}} \\
        \midrule
        \textbf{ONX~\cite{ONX}} & 8 nm & 4 MB & 512 CUDA Cores & 450 & 15 \\
        \textbf{RTX 3090~\cite{rtx3090}} & 8 nm & 80.25 MB & 5248 CUDA Cores & 628 & 352 \\
        \textbf{GauSPU~\cite{GauSPU}} & 12 nm & 560 KB & 128 REs/32 BEs & 30 & 9.4 \\
        \textbf{RTGS} & 28 nm & 197 KB & 16 REs/16 PEs & 28.41 & 8.11 \\
        \textbf{RTGS-12nm$^{1}$} & 12 nm & 197 KB & 16 REs/16 PEs & 6.49 & 4.63 \\
        \textbf{RTGS-8nm$^{1}$} & 8 nm & 197 KB & 16 REs/16 PEs & 2.40 & 3.76 \\
        \bottomrule
    \end{tabular}
    }
    \raggedright
    \footnotesize 
    $^{1}$ The 12nm and 8nm data are scaled form the \textit{DeepScaleTool} \cite{DST} with a voltage of 0.8V and a frequency of 500MHz.
    \label{tab:device_specs}
\vspace{-15pt}
\end{table}

\textbf{{\name} Simulation Setup.}
\underline{Simulation Method \& System Setup} \underline{Parameters:}
To evaluate the performance of {\name} when integrated with GPUs, we develop a cycle-accurate simulator based on GPGPU-Sim~\cite{GPGPU-sim}. 
We adopt a 500 MHz clock frequency based on an conservative modeling consideration since our hardware plug-in targets the 28 nm technology node.
The simulator is configured to closely reflect the ONX architecture. (1) For on-chip GPU configuration, we model 8 SMs, each with 32 threads per warp, 48KB of shared memory, and 128KB of L1 cache. A unified 2MB L2 cache is shared across all SMs.
(2) For off-chip memory, we configure a 128-bit LPDDR5 interface with a peak bandwidth of 104 GB/s. DRAM latency and energy consumption are simulated based on standard LPDDR5.

\underline{Simulator Test Trace Derivation:} We use GPGPU-Sim to model the interactions between SMs and RTGS. The key interactions include: (1) transferring 2D Gaussians from SMs to RTGS after sorting, and (2) returning Gaussian gradients from RTGS to SMs during backpropagation for pruning. To simulate the corresponding communication overhead, we extract memory access traces from the actual execution of 3DGS-SLAM on ONX edge GPU. These traces include information such as data volume and access patterns, and are used as input to GPGPU-Sim to enable accurate modeling of the communication overhead between SMs and RTGS.

\underline{System Simulation Validation:}
To validate simulation accuracy, we estimate power and runtime for preprocessing and sorting individually. On ONX edge GPU, preprocessing consumes 1.91 W on average and lasts for 0.92 ms, while sorting consumes 5.29 W and takes 2.55 ms. GPGPU-Sim reports 1.76 W and 0.83 ms for preprocessing, and 4.88 W and 2.42 ms for sorting. All relative errors in both runtime and power measurements are within 10\%, demonstrating that our GPGPU-Sim-based simulation setup provides accurate modeling of execution characteristics. Based on the validated simulator, our result shows that the DRAM bandwidth utilization is only 21.5\%, while the L2 cache utilization reaches 43.6\%, indicating that the memory traffic is more concentrated at the L2 level. Therefore, only small on-chip memory footprint s required.


\begin{table*}[!t]
\centering
\caption{Performance comparison of 3DGS-SLAM variants across four datasets.}
\vspace{-10pt}
\scriptsize
\renewcommand{\arraystretch}{1.2}
\setlength{\tabcolsep}{5pt}
\begin{tabular}{c|cccc|cccc|cccc|cccc}
\hline
\multirow{2}{*}{\textbf{Method}} 
& \multicolumn{4}{c|}{\textbf{TUM~\cite{TUM}}} 
& \multicolumn{4}{c|}{\textbf{Replica~\cite{Replica}}} 
& \multicolumn{4}{c|}{\textbf{ScanNet~\cite{Scannet}}} 
& \multicolumn{4}{c}{\textbf{ScanNet++~\cite{Scannet++}}} \\
& \makecell{ATE\\(cm)}& \makecell{PSNR\\(dB)} & \makecell{FPS\\} & \makecell{Mem\\(GB)} 
& \makecell{ATE\\(cm)} & \makecell{PSNR\\(dB)} & \makecell{FPS\\} & \makecell{Mem\\(GB)} 
& \makecell{ATE\\(cm)} & \makecell{PSNR\\(dB)} & \makecell{FPS\\} & \makecell{Mem\\(GB)} 
& \makecell{ATE\\(cm)} & \makecell{PSNR\\(dB)} & \makecell{FPS\\} & \makecell{Mem\\(GB)} \\
\hline
GS-SLAM~\cite{GSSLAM} & \cellcolor[HTML]{C8E6C9}3.7 & \cellcolor[HTML]{C8E6C9}15.93 & 3.3 & 8.3 & \cellcolor[HTML]{66BB6A}0.5 & \cellcolor[HTML]{C8E6C9}35.41 & 2.3 & 9.2 & \cellcolor[HTML]{C8E6C9}2.85 & \cellcolor[HTML]{C8E6C9}19.87 & 1.4 & 10.4 & \cellcolor[HTML]{C8E6C9}3.21 & \cellcolor[HTML]{C8E6C9}24.41 & 0.92 & 11.1 \\
Taming 3DGS+GS-SLAM & 6.7 & 14.31 & \cellcolor[HTML]{C8E6C9}4.7 & \cellcolor[HTML]{C8E6C9}4.2 & 3.2 & 30.3 & \cellcolor[HTML]{C8E6C9}3.2 & \cellcolor[HTML]{C8E6C9}4.6 & 5.6 & 14.3 & \cellcolor[HTML]{C8E6C9}1.9 & \cellcolor[HTML]{66BB6A}1.9 & 6.2 & 17.71 & \cellcolor[HTML]{C8E6C9}1.3 & \cellcolor[HTML]{C8E6C9}5.6 \\
Ours+GS-SLAM & \cellcolor[HTML]{66BB6A}3.4 & \cellcolor[HTML]{66BB6A}16.01 & \cellcolor[HTML]{66BB6A}12.1 & \cellcolor[HTML]{66BB6A}3.9 & \cellcolor[HTML]{C8E6C9}0.51 & \cellcolor[HTML]{66BB6A}35.44 & \cellcolor[HTML]{66BB6A}8.3 & \cellcolor[HTML]{66BB6A}4.3 & \cellcolor[HTML]{66BB6A}2.76 & \cellcolor[HTML]{66BB6A}21.75 & \cellcolor[HTML]{66BB6A}5.1 & \cellcolor[HTML]{C8E6C9}4.9 & \cellcolor[HTML]{66BB6A}3.19 & \cellcolor[HTML]{66BB6A}25.13 & \cellcolor[HTML]{66BB6A}3.3 & \cellcolor[HTML]{66BB6A}5.2 \\
\hline
MonoGS~\cite{MONOGS} & \cellcolor[HTML]{C8E6C9}1.47 & \cellcolor[HTML]{C8E6C9}25.82 & 1.8 & 13.1 & \cellcolor[HTML]{C8E6C9}0.32 & \cellcolor[HTML]{C8E6C9}38.94 & 1.2 & 13.5 & \cellcolor[HTML]{66BB6A}3.25 & \cellcolor[HTML]{C8E6C9}20.43 & 0.7 & 14.6 & \cellcolor[HTML]{C8E6C9}7.46 & \cellcolor[HTML]{66BB6A}23.79 & 0.6 & 15.0 \\
Taming 3DGS+MonoGS & 3.21 & 20.28 & \cellcolor[HTML]{C8E6C9}2.6 & \cellcolor[HTML]{C8E6C9}6.9 & 0.43 & 32.51 & \cellcolor[HTML]{C8E6C9}1.9 & \cellcolor[HTML]{C8E6C9}7.1 & 4.33 & 17.26 & \cellcolor[HTML]{C8E6C9}1.1 & \cellcolor[HTML]{C8E6C9}7.6 & 9.81 & 20.15 & \cellcolor[HTML]{C8E6C9}0.8 & \cellcolor[HTML]{C8E6C9}7.8 \\
Ours+MonoGS & \cellcolor[HTML]{66BB6A}1.41 & \cellcolor[HTML]{66BB6A}25.73 & \cellcolor[HTML]{66BB6A}4.7 & \cellcolor[HTML]{66BB6A}6.2 & \cellcolor[HTML]{66BB6A}0.29 & \cellcolor[HTML]{66BB6A}39.14 & \cellcolor[HTML]{66BB6A}3.6 & \cellcolor[HTML]{66BB6A}6.4 & \cellcolor[HTML]{C8E6C9}3.26 & \cellcolor[HTML]{66BB6A}20.44 & \cellcolor[HTML]{66BB6A}2.8 & \cellcolor[HTML]{66BB6A}6.1 & \cellcolor[HTML]{66BB6A}6.76 & \cellcolor[HTML]{C8E6C9}23.6 & \cellcolor[HTML]{66BB6A}1.6 & \cellcolor[HTML]{66BB6A}7.1 \\
\hline
Photo-SLAM~\cite{photoslam} & \cellcolor[HTML]{C8E6C9}2.61 & \cellcolor[HTML]{C8E6C9}20.12 & 8.1 & 4.3 & \cellcolor[HTML]{C8E6C9}0.64 & \cellcolor[HTML]{66BB6A}31.97 & 8.4 & 7.1 & \cellcolor[HTML]{C8E6C9}3.73 & \cellcolor[HTML]{C8E6C9}21.33 & 6.2 & \cellcolor[HTML]{C8E6C9}6.3 & \cellcolor[HTML]{C8E6C9}6.43 & \cellcolor[HTML]{C8E6C9}25.31 & 6.2 & \cellcolor[HTML]{C8E6C9}4.9 \\
Taming 3DGS+Photo-SLAM & 4.21 & 19.23 & \cellcolor[HTML]{C8E6C9}11.3 & \cellcolor[HTML]{C8E6C9}2.2 & 1.23 & 27.66 & \cellcolor[HTML]{66BB6A}11.1 & \cellcolor[HTML]{C8E6C9}2.6 & 4.1 & 20.33 & \cellcolor[HTML]{C8E6C9}8.8 & 8.8 & 6.99 & 23.12 & \cellcolor[HTML]{C8E6C9}8.9 & 6.4 \\
Ours+Photo-SLAM & \cellcolor[HTML]{66BB6A}2.33 & \cellcolor[HTML]{66BB6A}21.34 & \cellcolor[HTML]{66BB6A}12.96 & \cellcolor[HTML]{66BB6A}2.0 & \cellcolor[HTML]{66BB6A}0.61 & \cellcolor[HTML]{C8E6C9}31.9 & \cellcolor[HTML]{66BB6A}11.1 & \cellcolor[HTML]{66BB6A}2.3 & \cellcolor[HTML]{66BB6A}3.68 & \cellcolor[HTML]{66BB6A}22.45 & \cellcolor[HTML]{66BB6A}10.2 & \cellcolor[HTML]{66BB6A}2.2 & \cellcolor[HTML]{66BB6A}6.33 & \cellcolor[HTML]{66BB6A}26.54 & \cellcolor[HTML]{66BB6A}9.92 & \cellcolor[HTML]{66BB6A}2.3 \\
\hline
\end{tabular}
 \vspace{-1pt}
\label{tab:slam_units_fps}

\end{table*}

\textbf{Hardware Baselines.}
Our proposed {\name} hardware works as a general GPU plug-in module. Therefore, we choose three sets of hardware baselines: the \underline{base algorithm implementations on GPUs}, \underline{one GPU-based optimization technique}, i.e., DISTWAR~\cite{distwar}, to accelerate atomic gradient aggregations, and \underline{one GPU plug-in}, i.e., GauSPU~\cite{GauSPU}.
Specifically, since DISTWAR is open-sourced\footnote{https://github.com/Accelsnow/gaussian-splatting-distwar}, we integrate it with three keyframe-based algorithms, i.e., GS-SLAM~\cite{GSSLAM}, MonoGS~\cite{MONOGS}, and Photo-SLAM~\cite{photoslam}, on the ONX edge GPU. 
For GauSPU, which is developed on the NVIDIA GeForce RTX 3090 GPU~\cite{rtx3090}, we deploy our {\name} on the same GPU to ensure a fair comparison. GauSPU is not open source.


\begin{table}[b]
\vspace{-11pt}
    \centering
    \setlength{\tabcolsep}{2pt}
    \renewcommand{\arraystretch}{0.95}
    \caption{Performance comparison with GauSPU~\cite{GauSPU}, using SplaTAM~\cite{splatam} algorithm on RTX 3090 GPU~\cite{rtx3090}.}
    \vspace{-10pt}
    \resizebox{0.95\columnwidth}{!}{
  
    \begin{tabular}{c||c|c|c|c|c}
        \toprule
        \textbf{Method} & \makecell{ATE\\(cm) $\downarrow$} & \makecell{PSNR\\(dB) $\uparrow$} & \makecell{Tracking\\FPS $\uparrow$} & \makecell{Overall\\FPS $\uparrow$} & \makecell{Peak-Memory\\Usage (GB) $\downarrow$} \\
        \midrule
        SplaTAM~\cite{splatam} & 0.36 & 32.81 & 2.7 & 2.3 & 12.3 \\
        GauSPU + SplaTAM & \cellcolor[HTML]{C8E6C9}0.33 & \cellcolor[HTML]{66BB6A}34.00 & 14.6 & 11.4 & \cellcolor[HTML]{C8E6C9}7.3 \\
        Ours + SplaTAM & \cellcolor[HTML]{66BB6A}0.31 & \cellcolor[HTML]{C8E6C9}33.90 & \cellcolor[HTML]{66BB6A}22.6 & \cellcolor[HTML]{66BB6A}22.6 & \cellcolor[HTML]{66BB6A}5.9 \\
        \bottomrule
    \end{tabular}
    }
    \label{tab6}
\end{table}

\subsection{Evaluating {\name} Algorithm}
\label{sec:Exp_alg}
In this section, we evaluate the algorithm with four commonly-used metrics.  ATE measures the accuracy of camera trajectory reconstruction, PSNR reflects the fidelity of rendered images, Frames Per Second (FPS) indicates runtime performance, and Peak Memory Usage captures the maximum memory footprint.

\begin{figure}[t]
\vspace{-5pt}
    \centering
    \includegraphics[width=1.0\linewidth]{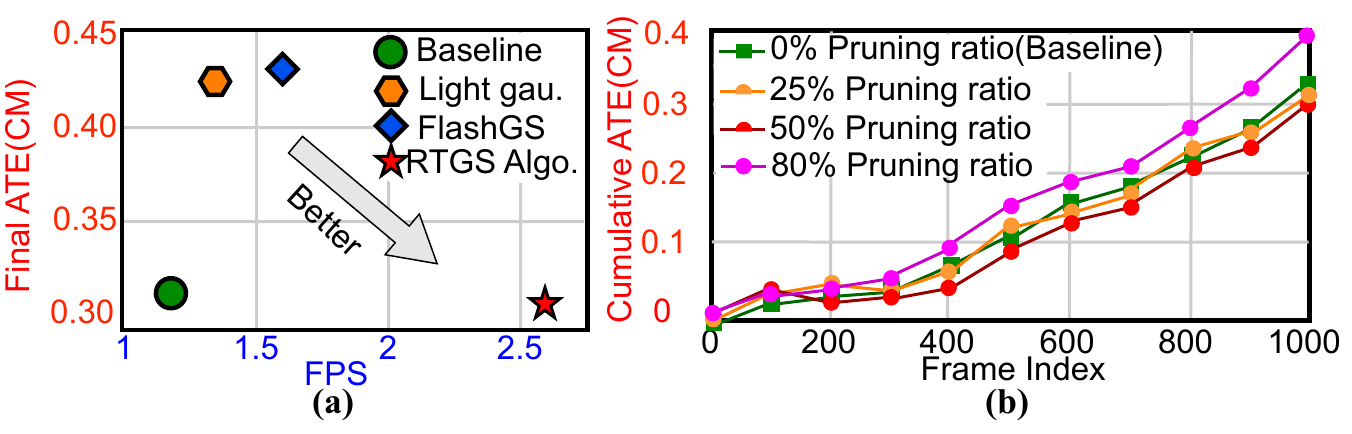}
    \vspace{-22pt}
    \caption {(a) Accuracy and efficiency trade-off analysis and (b) impact of adaptive pruning on long-term drift accumulation, using MonoGS~\cite{MONOGS} on the Replica~\cite{Replica} dataset.}
    \label{fig:rebuttal}
  \vspace{-15pt}
\end{figure}

\textbf{Benchmark with existing base algorithms and the Gaussian pruning technique.}
Tab.~\ref{tab:slam_units_fps} summarizes the performance of three keyframe-based 3DGS-SLAM base algorithms, along with their variants enhanced by Taming 3DGS~\cite{Taming} pruning and our proposed {\name}. 
By eliminating redundant Gaussians and pixels, our {\name} achieves 2.5$\times$ - 3.6$\times$ speedup with less than 5\% degradation in ATE and PSNR.
In contrast, Taming 3DGS requires thousands of iterations for pruning to converge, making it unsuitable for 3DGS-SLAM, which typically runs within 100 iterations.

\label{sec:alg_exp}

\textbf{Benchmark with more precise algorithms.} 
As shown in Fig.~\ref{fig:rebuttal}(a), our method achieves significantly higher FPS while maintaining comparable ATE and PSNR ccuracy comparable to the baseline. The FPS improvement comes from that our pruning strategy does not introduce additional overhead to the pipeline. In contrast, the compute-intensive importance evaluation used by LightGaussian and FlashGS adds extra operations that increase runtime. Our accuracy preservation is attributed to a pose-aware pruning strategy, where we selectively remove Gaussians that have minimal influence on camera pose updates. This ensures that our method reduces redundancy while preserving tracking performance.

\textbf{Benchmark with sparse sampling technique.}
Tab. \ref{tab6} compares our proposed {\name} with the sparse sampling technique GauSPU\cite{GauSPU}. Unlike GauSPU, which requires a customized GPU hardware plug-in to achieve FPS gains, our method delivers 22.6 FPS solely through algorithmic optimizations on RTX 3090 GPUs. This demonstrates that {\name} achieves comparable or better quality with significantly higher runtime performance.

\textbf{Ablation study on tracking long-term stability.} As shown in Fig.~\ref{fig:rebuttal}(b), our method maintains similar ATE growth trends as the unpruned one when pruning ratio $\leq$ 50\%. In some scenes, pruning even reduces drift by removing noisy Gaussians. However, at 60\% pruning, ATE rises sharply early on, as important Gaussians are mistakenly removed, leading to accumulated pose errors. This shows the need for a conservative pruning cap. Drift may be corrected via loop closure, and future work could explore longer sequences.

\begin{figure}[t]
\vspace{-10pt}
    \centering
    \includegraphics[width=1.0\linewidth]{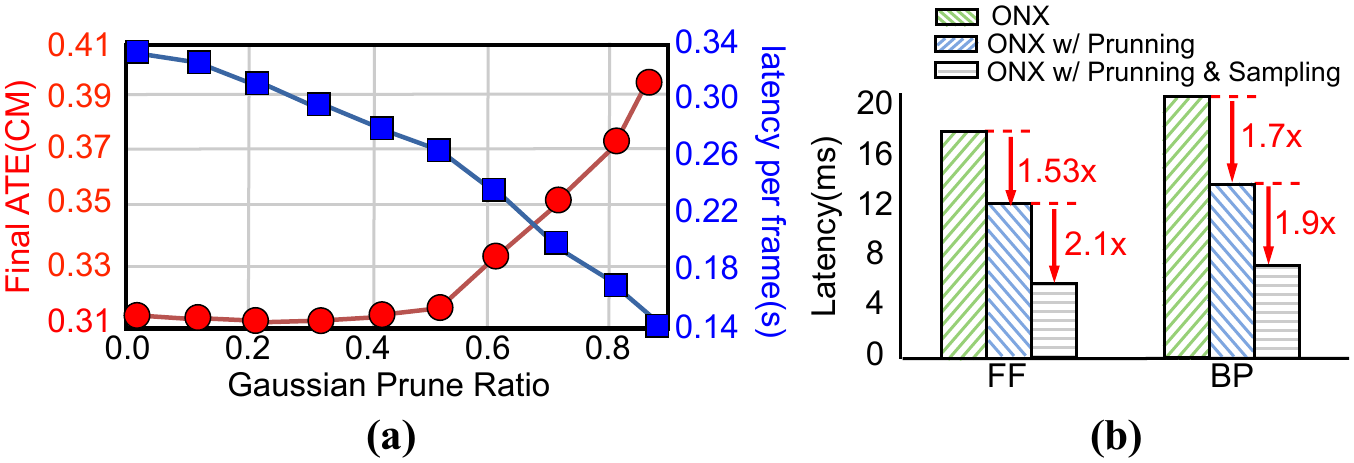}
    \vspace{-25pt}
    \caption {(a) Ablation study of Gaussian pruning ratio and (b) performance breakdown of {\name} algorithm techniques, using MonoGS~\cite{MONOGS} on the Replica~\cite{Replica} dataset. }
    \label{fig:ratio}
\vspace{-15pt}
\end{figure}

\textbf{Impact of pruning ratio.} 
As shown in Fig.~\ref{fig:ratio}(a), pruning ratios $\geq$ 50\% cause a sharp ATE increase, degrading accuracy. We therefore cap the pruning ratio at 50\% to balance performance and quality.

\textbf{Speedup breakdown.} 
Fig.~\ref{fig:ratio}(b) shows that adaptive pruning accelerates FF by 1.53$\times$ and BP by 1.7$\times$, while dynamic downsampling improves them by 2.1$\times$ and 1.9$\times$, respectively, confirming the effectiveness of both techniques.

\begin{figure*}[t]
    \centering
    \includegraphics[width=1\linewidth]{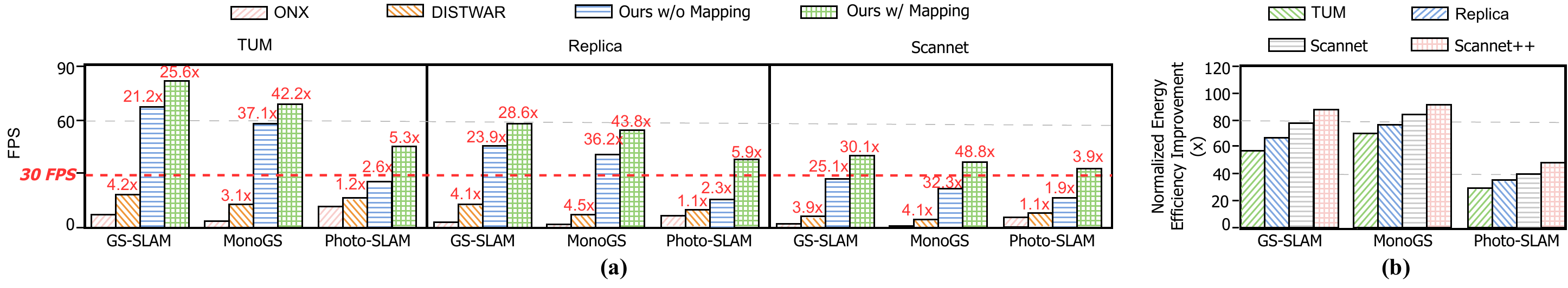}
    \vspace{-24pt}
    \caption{The FPS on the proposed {\name} and the baseline GPU. (a) Comparison of FPS across four baseline algorithms on three datasets using four configurations: ONX edge GPU~\cite{ONX}, {\name} with tracking acceleration only, and {\name} with both tracking and mapping acceleration.  (b) Improvement in energy efficiency across the three baseline algorithms on four datasets. }
    \label{fig:fps comparison}
\end{figure*}

\begin{figure*}[t]
  \centering
  \begin{minipage}{0.49\textwidth}
    \centering
    \includegraphics[width=\linewidth]{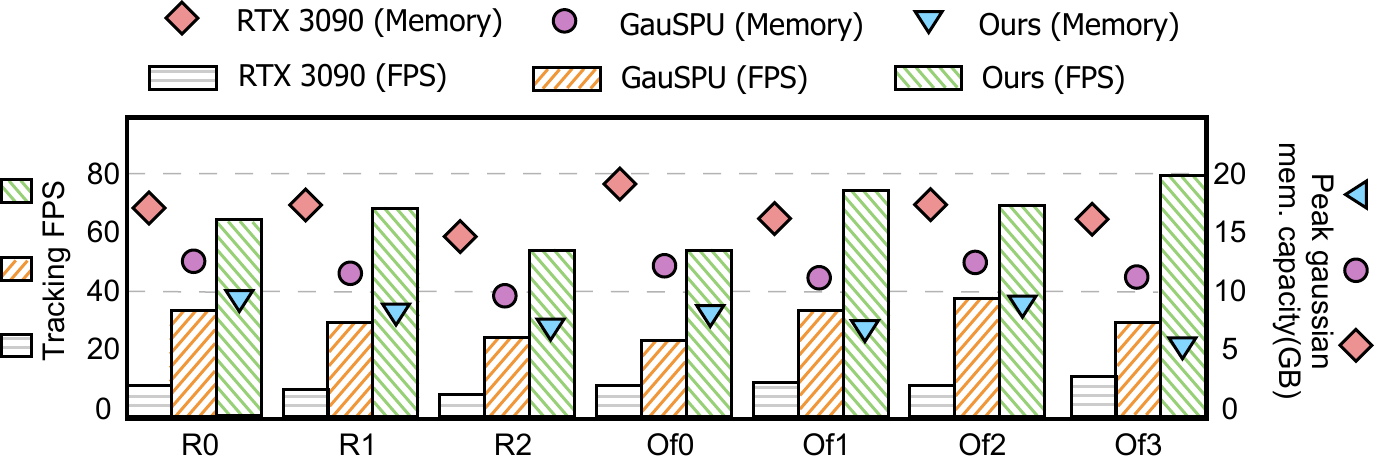}
    \vspace{-22pt}
    \caption{Comparison of RTX 3090~\cite{rtx3090}, GauSPU~\cite{GauSPU}, and proposed RTGS, using SplaTAM~\cite{splatam} algorithm on Replica~\cite{Replica} dataset: (a) tracking FPS and (b) memory efficiency.}
    \label{fig:GauSPU_compare}
  \end{minipage}
  \hfill
  \begin{minipage}{0.49\textwidth}
    \centering
    \includegraphics[width=\linewidth]{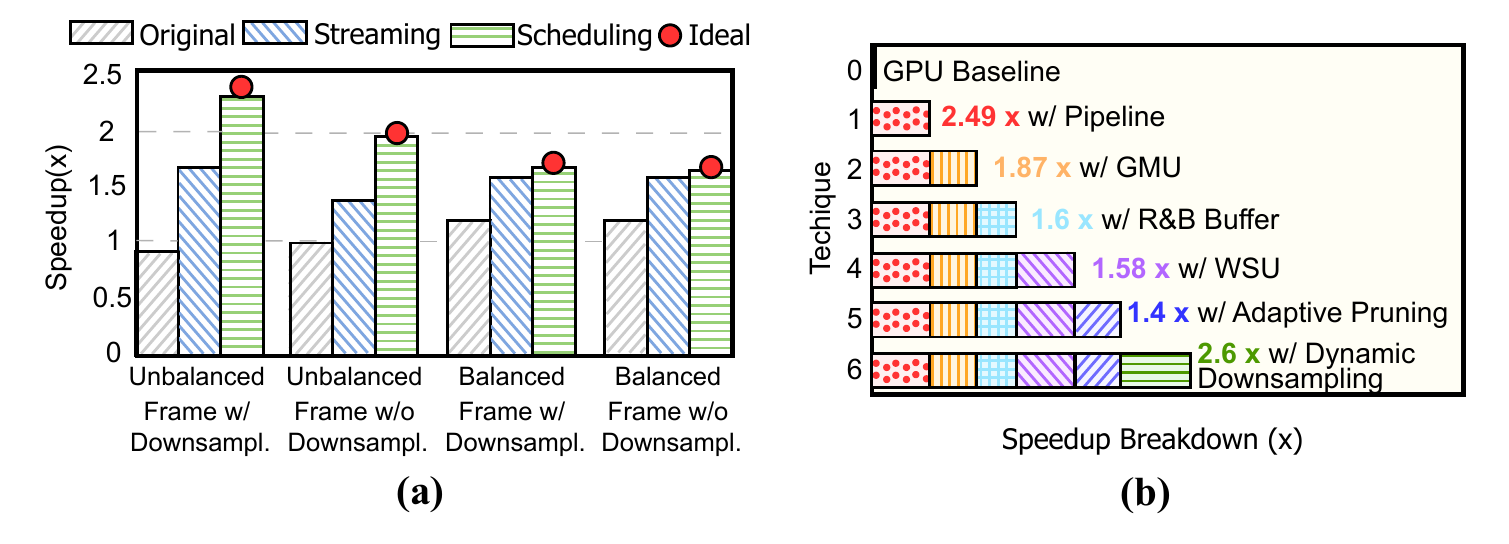}
    \vspace{-25pt}
    \caption{Ablation study of the performance breakdown for (a) two techniques for mitigating workload imbalance and (b) all \name~techniques on Replica~\cite{Replica} with MonoGS~\cite{MONOGS} baseline.}
    \label{fig:breakdown}
  \end{minipage}
\vspace{-10pt}
\end{figure*}




\subsection{Evaluating RTGS Architecture}
\label{sec:overall_exp}

\textbf{Benchmark with existing base algorithm and the GPU-based atomic operation acceleration technique.} 
\underline{Throughput speedup:} 
Fig.~\ref{fig:fps comparison}(a) presents a comparison of the end-to-end system FPS of the {\name}-enhanced ONX edge GPU against the base algorithm implementations on the ONX edge GPU, along with their variants using DISTWAR's GPU-based atomic operation acceleration technique. 
First, across the three datasets and three base algorithms, our {\name} consistently achieves real-time throughput performance, i.e., $\geq$30 FPS, while DISTWAR fails to reach real-time performance, demonstrating the necessity of multi-level redundancy reduction. 
Second, accelerating only the tracking stage still falls short of 30 FPS on large datasets such as ScanNet, emphasizing the importance of generalizable techniques for both tracking and mapping stages.
\underline{Energy efficiency improvement:}
Fig.~\ref{fig:fps comparison}(b) shows the overall energy efficiency improvements achieved by {\name}. Across the four datasets, TUM, Replica, ScanNet, and ScanNet++, {\name} achieves average energy efficiency improvements (measured as energy per frame) of 32.7$\times$, 56.9$\times$, 73.0$\times$, and 69.4$\times$, respectively.

\textbf{Benchmark with the GPU plug-in.} 
\underline{Area and power:} Tab.~\ref{tab:device_specs} shows the hardware specifications of {\name}, GauSPU~\cite{GauSPU}, and two GPUs. Thanks to its algorithm optimizations, {\name} uses less on-chip SRAM and fewer compute cores compared to GauSPU, resulting in a smaller area and lower power consumption for edge deployment.
\underline{Tracking throughput and memory efficiency:} By plugging {\name} into the RTX 3090, as done with GauSPU, Fig.~\ref{fig:GauSPU_compare} shows that {\name} achieves higher tracking FPS and reduces peak Gaussian memory capacity compared to GauSPU. On average, our approach yields a 2.3$\times$ improvement in FPS and a 1.3$\times$ reduction in peak memory consumption. These results show the effectiveness of {\name} in reducing multi-level redundancies and achieving speedup.

Our memory efficiency improvement comes from adaptive pruning and architectural optimizations, which reduce redundant Gaussians and minimize overhead from data reorganization and access, ensuring efficient operation under dynamic workloads.

\textbf{Speedup breakdown among two techniques for mitigating intra- and inter-subtile workload imbalance.}
Fig.~\ref{fig:breakdown}(a) shows the speedup from subtile-level streaming alone, and further gains with pixel-level pairwise scheduling. Their combination approaches the ideal speedup bound, highlighting the importance of integrating both techniques for effective workload balancing.

\textbf{Overall speedup breakdown.} 
We evaluate the proposed techniques on MonoGS with fr1/desk scene in TUM, with detailed speedup decomposition in Fig.\ref{fig:breakdown}(b): (1) the design of RE and PE shows a 2.49$\times$ improvement due to pipelined execution; (2) on the \textbf{step level}, Gradient Merging Unit further improves the FPS by 1.87$\times$ 
; (3) the performance improves by 1.6 $\times$ due to R\&B buffer reuse; (4) integration with Workload Scheduling Unit achieves a 1.58$\times$ speedup by balancing workload; (5) on the \textbf{iteration level}, the adoption of Adaptive Gaussian Pruning accelerates the execution speed of non-keyframes; (6) on the \textbf{frame level}, Dynamic Downsampling 
further accelerates by 2.60$\times$.


\section{Related Work}
\label{sec:related_work}


\textbf{3DGS Acceleration:} With 3DGS~\cite{Kerbl3DGS} achieving high speed and quality, recent efforts have explored software~\cite{EAGLES,LightGaussian,compress_gaussian} and hardware~\cite{gscore,gsarch,lin2025metasapiens,distwar,GauSPU} optimizations. LightGaussian~\cite{LightGaussian} reduces memory via distillation algorithmically. GSArch~\cite{gsarch} and DISTWAR~\cite{distwar} target the atomic bottlenecks by respectively introducing gradient filtering and warp-level aggregation. Compared to prior works, our design exploits multi-level redundancy across the 3DGS pipeline and achieves higher performance with minimal overhead.
\section{Conclusion and Future Work}
\label{sec:con}

In this paper, we propose {\name}, a GPU-integrated accelerator for 3DGS-based SLAM that achieves over 30 FPS real-time performance by reducing multi-level redundancies through algorithm-hardware co-design with minimal overhead. 
Beyond 3DGS-SLAM, our co-design techniques are also applicable to differentiable rendering systems such as NvDiffRec~\cite{NvDiffRec} and Pulsar~\cite{Pulsar}. 
Our strategies can be integrated into these systems to alleviate workload imbalance and improve overall throughput.

\begin{acks}
This work was partially supported by Cisco Gift Funds and an Amazon Research Award (PI: Prof.\ Tianlong Chen).
\end{acks}
\clearpage

\bibliographystyle{ACM-Reference-Format}
\bibliography{references}
\end{document}